%% file: draft78.tex
\documentclass[aps,prd,twocolumn,nofootinbib,showpacs,superscriptaddress,showkeys]{revtex4-1}

\usepackage{graphicx}  
\usepackage{bm}        
\usepackage{amssymb}
\usepackage{amsmath}

\usepackage{pstricks}
\usepackage{multirow}

\fontencoding{T1}
\fontfamily{pag}
\selectfont

\newcommand{\be}{\begin{equation}}
\newcommand{\ee}{\end{equation}}
\newcommand{\beqa}{\begin{eqnarray}}
\newcommand{\eeqa}{\end{eqnarray}}

\newcommand{\la}{\langle}
\newcommand{\ra}{\rangle}
\newcommand{\ph}{{ \phi}}

\newcommand{\Pt}{{ P_{h\perp} }}
\newcommand{\uu}{{\scriptscriptstyle UU}}
\newcommand{\uut}{{\scriptscriptstyle UU,T}}
\newcommand{\uul}{{\scriptscriptstyle UU,L}}

\newcommand{\cs}{\la {\cos\ph} \ra}
\newcommand{\cst}{\la {\cos2\ph} \ra}
\newcommand{\csN}{\la {\cos n\ph} \ra}

\newcommand{\csm}{\cos\ph}
\newcommand{\cstm}{\cos2\ph}
\newcommand{\csNm}{\cos n\ph}

\newcommand{\BM}{Boer--Mulders}
\newcommand{\BMC}{Boer--Mulders--Collins}
\newcommand{\C}{Cahn}
\newcommand{\df}{distribution function}
\newcommand{\ff}{fragmentation function}
\newcommand{\hermes}{{\sc{Hermes}}}
\newcommand{\desy}{{\sc{Desy}}}
\newcommand{\hera}{{\sc Hera}}
\newcommand{\clas}{{\sc Clas}}
\newcommand{\compass}{{\sc Compass}}

\newcommand{\pythia}{{\sc Pythia6}}
\newcommand{\jetset}{{\sc Jetset}}
\newcommand{\radgen}{{\sc Radgen}}
\newcommand{\geant}{{\sc Geant}}
\newcommand{\cteq}{{\sc Cteq}}
\newcommand{\MC}{{Monte Carlo}}
\newcommand{\lepto}{{\sc LEPTO}}

\graphicspath{{./}{./Figures/}{./Figures/Multi/}}

\begin{document}

\title{Azimuthal distributions of charged hadrons, 
pions, and kaons produced in deep-inelastic scattering 
off unpolarized protons and deuterons }

\input{rec-cosphi_final.tex}

\begin{abstract}
The azimuthal $\csm$ and $\cstm$ modulations of the distribution of hadrons produced in 
unpolarized semi-inclusive deep-inelastic scattering of electrons and 
positrons off hydrogen and deuterium targets have been measured in the \hermes\ experiment.
For the first time these modulations were determined 
in a four-dimensional kinematic space for positively and negatively charged pions and kaons separately, 
as well as for unidentified hadrons.
These azimuthal dependences are sensitive to the transverse motion and polarization of the quarks 
within the nucleon via, e.g., the Cahn, Boer-Mulders and Collins effects.
\end{abstract}

\pacs{13.88.+e, 13.60.-r}

\keywords{Semi-inclusive deep-inelastic scattering, azimuthal modulations, 
intrinsic transverse momentum and spin.}

\maketitle

\input{intro}

\input{analysis}
\input{results}
\appendix
\input{appendix}

\bibliography{cosphi}

\end{document}

%% file: rec-cosphi_final.tex
\def\groupargonne{\affiliation{Physics Division, Argonne National Laboratory, Argonne, Illinois 60439-4843, USA}}
\def\groupbari{\affiliation{Istituto Nazionale di Fisica Nucleare, Sezione di Bari, 70124 Bari, Italy}}
\def\groupbeijing{\affiliation{School of Physics, Peking University, Beijing 100871, China}}
\def\groupbilbao{\affiliation{Department of Theoretical Physics, University of the Basque Country UPV/EHU, 48080 Bilbao, Spain and IKERBASQUE, Basque Foundation for Science, 48011 Bilbao, Spain}}
\def\groupcolorado{\affiliation{Nuclear Physics Laboratory, University of Colorado, Boulder, Colorado 80309-0390, USA}}
\def\groupdesy{\affiliation{DESY, 22603 Hamburg, Germany}}
\def\groupzeuthen{\affiliation{DESY, 15738 Zeuthen, Germany}}
\def\groupdubna{\affiliation{Joint Institute for Nuclear Research, 141980 Dubna, Russia}}
\def\grouperlangen{\affiliation{Physikalisches Institut, Universit\"at Erlangen-N\"urnberg, 91058 Erlangen, Germany}}
\def\groupferrara{\affiliation{Istituto Nazionale di Fisica Nucleare, Sezione di Ferrara and Dipartimento di Fisica, Universit\`a di Ferrara, 44100 Ferrara, Italy}}
\def\groupfrascati{\affiliation{Istituto Nazionale di Fisica Nucleare, Laboratori Nazionali di Frascati, 00044 Frascati, Italy}}
\def\groupgent{\affiliation{Department of Physics and Astronomy, Ghent University, 9000 Gent, Belgium}}
\def\groupgiessen{\affiliation{II. Physikalisches Institut, Justus-Liebig-Universit\"at Gie{\ss}en, 35392 Gie{\ss}en, Germany}}
\def\groupglasgow{\affiliation{SUPA, School of Physics and Astronomy, University of Glasgow, Glasgow G12 8QQ, United Kingdom}}
\def\groupillinois{\affiliation{Department of Physics, University of Illinois, Urbana, Illinois 61801-3080, USA}}
\def\groupmichigan{\affiliation{Randall Laboratory of Physics, University of Michigan, Ann Arbor, Michigan 48109-1040, USA }}
\def\groupmoscow{\affiliation{Lebedev Physical Institute, 117924 Moscow, Russia}}
\def\groupnikhef{\affiliation{National Institute for Subatomic Physics (Nikhef), 1009 DB Amsterdam, The Netherlands}}
\def\groupstpetersburg{\affiliation{B.~P.~Konstantinov Petersburg Nuclear Physics Institute, Gatchina, 188300 Leningrad Region, Russia}}
\def\groupprotvino{\affiliation{Institute for High Energy Physics, Protvino, 142281 Moscow Region, Russia}}
\def\groupregensburg{\affiliation{Institut f\"ur Theoretische Physik, Universit\"at Regensburg, 93040 Regensburg, Germany}}
\def\grouprome{\affiliation{Istituto Nazionale di Fisica Nucleare, Sezione di Roma, Gruppo Collegato Sanit\`a and Istituto Superiore di Sanit\`a, 00161 Roma, Italy}}
\def\grouptriumf{\affiliation{TRIUMF, Vancouver, British Columbia V6T 2A3, Canada}}
\def\grouptokyo{\affiliation{Department of Physics, Tokyo Institute of Technology, Tokyo 152, Japan}}
\def\groupamsterdam{\affiliation{Department of Physics and Astronomy, VU University, 1081 HV Amsterdam, The Netherlands}}
\def\groupwarsaw{\affiliation{National Centre for Nuclear Research, 00-689 Warsaw, Poland}}
\def\groupyerevan{\affiliation{Yerevan Physics Institute, 375036 Yerevan, Armenia}}
\def\groupnone{\noaffiliation}

\groupargonne
\groupbari
\groupbeijing
\groupbilbao
\groupcolorado
\groupdesy
\groupzeuthen
\groupdubna
\grouperlangen
\groupferrara
\groupfrascati
\groupgent
\groupgiessen
\groupglasgow
\groupillinois
\groupmichigan
\groupmoscow
\groupnikhef
\groupstpetersburg
\groupprotvino
\groupregensburg
\grouprome
\grouptriumf
\grouptokyo
\groupamsterdam
\groupwarsaw
\groupyerevan


\author{A.~Airapetian}  \groupgiessen \groupmichigan
\author{N.~Akopov}  \groupyerevan
\author{Z.~Akopov}  \groupdesy
\author{E.C.~Aschenauer} \thanks{Now at: Brookhaven National Laboratory, Upton, New York 11772-5000, USA}\groupzeuthen
\author{W.~Augustyniak}  \groupwarsaw
\author{R.~Avakian}  \groupyerevan
\author{A.~Avetissian}  \groupyerevan
\author{E.~Avetisyan}  \groupdesy
\author{S.~Belostotski}  \groupstpetersburg
\author{H.P.~Blok}  \groupnikhef \groupamsterdam
% ...

\collaboration{The \hermes\ Collaboration} \noaffiliation

%% file: intro.tex
\section{Introduction}\label{sec:intro}

Since the late 1960s the quark-parton model~\cite{Bjorken:1969ja,Feynman:1969ej}
has been used to describe the structure of the nucleon in terms of fundamental constituents.
Their behavior inside the nucleon was parametrized in terms of parton distribution functions (PDFs).
Historically PDFs depended only on the fractional quark momentum
longitudinal to the nucleon direction of motion, $x$,
and on the scale at which the distributions were probed,  $Q^2$,
while transverse degrees of freedom were neglected.
These PDFs have provided a good description of processes in which transverse spin and momentum are 
integrated over~\cite{deFlorian:2009vb, Aaron:2009wt, Pumplin:2002vw}. 
However, transverse degrees of freedom are not a priori negligible and are needed for a complete description of the nucleon.
To account for transverse motion the PDFs have been generalized
to transverse-momentum-dependent PDFs, known also as 
TMDs~\cite{Mulders:1995dh,Collins:1981uk,Boer:1997nt,Ji:2004xq,Ji:2004wu,Collins:book}.

Already in the early days of the parton model it was realized that the inclusion of quark intrinsic transverse momentum, $p_T$, leads to modifications of 
the cross sections in high-energy reactions involving hadrons in the initial state, e.g., in lepton-nucleon deep-inelastic scattering (DIS)~\cite{Feynman72,Ravndal73}. 
In particular, in semi-inclusive DIS transverse momenta give rise to azimuthal dependences of the distribution of the
produced hadrons about the direction of the virtual photon~\cite{Ravndal73,Kingsley74,Kotzinian:1994dv,Tangerman:1995hw,Mulders:1995dh,Boer:1997nt}.
In 1978, R.~Cahn discussed the emergence of cosine modulations in the semi-inclusive DIS
cross section in the presence of non-vanishing transverse parton momentum using simple kinematic considerations (\C\ effect)~\cite{Cahn:1978se,Cahn:1989yf}.
Once included, interplay between
the parton transverse momentum and the partons's and nucleon's 
spins can generate further azimuthal 
asymmetries, as, e.g.,  in the \BM\ effect.
The \BM\ mechanism was introduced for the first time
in 1997~\cite{Boer:1997nt} in relation to 
naive-T-odd effects\footnote{A 
naive-T-odd transformation is defined to 
be T-odd in the usual sense except without the 
interchange of initial and final states.}~\cite{Sivers:1990cc,Sivers:1991fh}.
For a long time naive-T-odd effects were believed 
to vanish due to time-reversal invariance~\cite{Collins:1993kk}.
Recently, it was shown that final and initial-state interactions
can produce naive-T-odd effects without violating 
T-invariance~\cite{Brodsky:2002cx,Collins:2002kn,Belitsky:2002sm}.
Because it involves only the parton spin and not the nucleon spin,
the \BM\ mechanism is a good example of how spin-related effects may
play an important role, even in unpolarized reactions.
Measurements of these novel correlations provide insights into the so far poorly explored
partonic transverse degrees of freedom and can be used to gather information,
in a model-dependent way, about the elusive parton orbital motion.

\begin{figure}
\begin{center}
\parbox{1\linewidth}
{\includegraphics[width=0.99\linewidth]{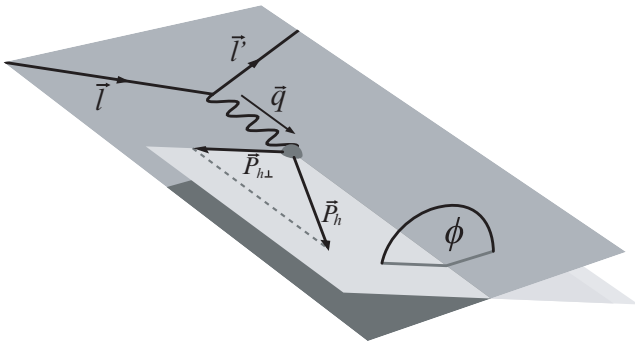}}
\caption{Depiction of the azimuthal angle $\ph$ between the scattering plane, 
spanned by the three-momenta ($\vec{l}$, $\vec{l'}$) of incoming and outgoing leptons
and the hadron plane, defined by the respective three-momenta of the virtual photon
and the produced hadron, $\vec{q}$ and $\vec{P_h}$, defined
according to the Trento convention~\cite{Bacchetta:2004jz}.}
\label{fig:evento}
\end{center}
\end{figure}

In DIS the structure of the nucleon is 
probed by the interaction of a high-energy lepton ($l$) with a target nucleon ($N$)  
via the exchange of electroweak bosons. In the kinematic region accessed at \hermes\,
it is a good approximation to consider only the exchange of a single photon (Born approximation)~\cite{Airapetian:2009wj}.
In semi-inclusive DIS measurements, at least one of the hadrons 
($h$) produced in the collision is detected in coincidence with the scattered lepton ($l'$):
\be
l\,+\,N\, \rightarrow \, l'\, + \,h\,\,+\,X,
\ee
where $X$ represents the remaining, unobserved, final state.
The polarization-averaged semi-inclusive DIS cross section can be written 
in a model-independent way by means of four structure functions~\cite{ChengZee72,Bacchetta:2006tn}:
\begin{align}
 d&{\sigma}_\uu \equiv \frac{{\mathrm d}^5\sigma_\uu}{{\mathrm d}x\,{\mathrm d}y\,{\mathrm d}z\,{\mathrm d}P^2_{h\perp}\,{\mathrm d}\ph} =\notag  \\ 
\quad & 2\pi\frac{\alpha^2}{xyQ^2}\frac{y^2}{2(1-\epsilon)}\left(1+\frac{\gamma^2}{2x}\right) \{ F_\uut+ \epsilon\, F_\uul\notag  \\ 
& + \sqrt{2\epsilon(1+\epsilon)} \, F_\uu^{\csm}\csm +\epsilon \, F_\uu^{\cstm}\cstm\}.
\label{eq:noncol_csec}
\end{align}
Here,
$-Q^2$ is the squared four-momentum carried by the virtual photon.
In the target rest frame, $y$ is the fraction of the beam energy carried by the virtual photon and
$z$ is the fraction of the virtual photon energy carried by the produced hadron.
The hadron momentum component transverse to the virtual photon direction is denoted $\Pt$, and 
$\ph$ is the azimuthal angle of the hadron production plane around the 
virtual photon direction with respect to the lepton scattering plane 
(see Fig.~\ref{fig:evento}). 
The quantity $\alpha$ is the electromagnetic coupling constant, and 
$\gamma=2Mx/Q$ with $M$ the proton mass. 
The structure functions $F_\uut,\, F_\uul ,\,F_\uu^{\csm},\,F_\uu^{\cstm}$ depend on $x$, $Q^2$, $z$ 
and $\Pt$; the subscript {$\scriptstyle UU$} stands for unpolarized beam and target, 
while $\scriptstyle T$ ($\scriptstyle L$) indicates transverse (longitudinal) 
polarization of the virtual photon, and $\epsilon$ is the ratio of
longitudinal to transverse photon flux.

Besides intrinsic transverse parton momentum, perturbative-QCD (pQCD) effects, like gluon
radiation, will also lead to azimuthal dependences in the semi-inclusive
DIS cross section~\cite{GeorgiPolitzer78, Mendez78}. However, they contribute mainly at large values of
$\Pt$, and are next-to-leading order in the strong coupling constant.
For hadron transverse momenta that are small compared to the hard scale $Q$
($\Pt \ll Q$), TMD factorization~\cite{Collins:1981uk,Ji:2004xq,Ji:2004wu}
allows for an expansion  of the structure functions in powers of 
$1/Q$ and expresses them 
in terms of convolutions 
of transverse-momentum dependent distribution and fragmentation functions.

The TMDs parametrize the nucleon structure and
the fragmentation functions describe how the struck parton evolves 
into the observed hadronic final state. 
For simplicity, in the following the weak $Q^2$ dependence of the TMDs
is not explicitly written.
Each convolution can be classified according to the suppression, in 
powers of $1/Q$, at which they contribute to the structure function.
Not all contributions from all possible suppression levels have been calculated yet.
In this work primarily contributions up to a suppression of $(1/Q)$ will be considered, but, 
considering the low average $Q^2$ attainable at \hermes,
contributions suppressed as $(1/Q)^2$ or higher
may be not negligible.\\

The structure function related to the $\cstm$ amplitude, $F_\uu^{\cstm}$, 
receives a single unsuppressed contribution:
\be
\begin{split}
 F_\uu^{\cstm}\propto -\sum_q\Bigl[&h_1^{\perp,q}(x,p^2_T) \otimes_{\mathcal W_1} H_1^{\perp,q}(z,k^2_T)\Bigr].
\label{eq:Twist2}
\end{split}
\ee
Additional contributions are present only at a suppression of $\propto(1/Q)^2$,
or higher. The sum symbol, $\sum_q$, stands for a quark-charge-squared weighted sum over quark flavors.
The symbol $\otimes_{\mathcal W_1}$ represents a weighted\footnote{The weights $\mathcal W_1$ (Eq.~\ref{eq:Twist2}), 
$\mathcal W_2$ (Eq.~\ref{eq:Twist4}), and $\mathcal W_3$ and  $\mathcal W_4$ 
(Eq.~\ref{eq:Twist3}) are kinematic factors depending on $p_T$ and $k_T$;
for their complete expressions see Ref.~\cite{Bacchetta:2006tn}.}
convolution integral over the intrinsic momentum $p_T$ and over $k_T$, 
the momentum transverse to the struck quark direction that
the hadron acquires during the fragmentation process. 

The contribution shown in Eq.~\ref{eq:Twist2} is called the \BMC\ effect (also often referred to as simply the \BM\ effect). It
involves the \BM\ \df\ $h_1^{\perp,q}(x,p_T^2)$~\cite{Boer:1997nt},
which describes the correlation between the transverse polarization and transverse momentum of quarks in an unpolarized nucleon,
and the Collins \ff\ $H_1^{\perp,q}(z,k^2_T)$~\cite{Collins:1993kk},
which describes the probability for a transversely polarized quark to
fragment to an unpolarized hadron with a certain transverse-momentum direction;
both these functions are chiral-odd. As hard QED and hard QCD interactions
preserve chirality, two chiral-odd functions need to
appear in conjunction, in order to have a chiral-even
observable.

As discussed above, naive-T-odd observables can be non-zero in
conjunction with final- or initial-state interactions,
which are reflected by the presence of non-trivial gauge links in the definition
of the TMDs~\cite{Collins:2002kn,Belitsky:2002sm}.
This  gauge link leads to the direct QCD prediction
that naive-T-odd distribution functions must have opposite signs
in semi-inclusive DIS and Drell--Yan reactions~\cite{Collins:2002kn}.
To date, this sign change has not yet been confirmed.

There are no contributions to $F_\uu^{\cstm}$ at a suppression $1/Q$. 
Not all contributions beyond a suppression of $1/Q$
have been calculated, however a term 
\be
\propto \Big(\frac{M}{Q}\Big)^2 \sum_q \Bigl[f_1^q(x,p^2_T) \otimes_{\mathcal W_2} D_1^q(z,k^2_T)\Bigr],
\label{eq:Twist4}
\ee
arising from the \C\ effect~\cite{Cahn:1978se,Cahn:1989yf} is expected. 
The \C\ effect has recently received increasing attention, 
as it can provide information about the average transverse momentum of unpolarized quarks in unpolarized hadrons.
It involves the convolution over transverse momenta of the spin-averaged
distribution and fragmentation functions, $f_1^q(x,p_T^2)$ and $D_1^q(z,k^2_T)$, 
respectively. Their transverse-momentum-integrated correspondents,
$f_1^q(x)$ and $D_1^q(z)$, are well known~\cite{Gluck:1998xa,deFlorian:2007aj}. 
However, their $p_T$- and $k_T$-dependences are poorly constrained by measurements,
and thus the convolution integrals in Eq.~\ref{eq:Twist4} 
can be estimated only approximately. 
Moreover, the average intrinsic transverse momentum $\la p_T\ra$ 
may depend on the parton flavor; thus a flavor-dependent 
measure of the \C\ effect, via, e.g., semi-inclusive DIS of identified hadrons,
is highly desirable.

The first non-zero contributions to the structure function $F_\uu^{\csm}$ in Eq.~\ref{eq:noncol_csec},
which is related to a $\csm$ amplitude,
are suppressed as $1/Q$;
subsequent contributions are suppressed as $(1/Q)^3$.
Among the various contributions suppressed as $1/Q$, several involve either a distribution or fragmentation
function that relates to quark-gluon-quark correlations, and hence is
interaction dependent and has no probabilistic interpretation.
In the Wandzura--Wilczeck approximation~\cite{Wandzura:1977qf}
all these terms are neglected, and only two contributions are considered: 
\be
\begin{split}
 F_\uu^{\csm}\simeq & -\frac{M}{Q}\sum_q \Bigl[h_1^{\perp,q}(x,p^2_T) \otimes_{\mathcal W_3} H_1^{\perp,q}(z,k^2_T)\Bigr]\\
& -\frac{M}{Q} \sum_q\Bigl[f_1^q(x,p^2_T) \otimes_{\mathcal W_4} D_1^q(z,k^2_T)\Bigr].
\label{eq:Twist3}
\end{split}
\ee
In the first line of Eq.~\ref{eq:Twist3} the \BMC\ effect is recognizable,
while, in the second line, the \C\ effect is present.

Only a few measurements of $\cstm$ and $\csm$ amplitudes in semi-inclusive DIS experiments
have been published over the past $30$ 
years~\cite{Aubert:1983cz,Arneodo:1986cf,Breitweg:2000qh,Adams:1993hs}.
Most measurements averaged over any possible flavor dependence
as they refer to hadrons without type nor charge distinction, 
and only hydrogen target~\cite{Aubert:1983cz,Arneodo:1986cf,Breitweg:2000qh}
or hydrogen and deuterium targets combined together~\cite{Adams:1993hs} were available.
Recently, the \clas\ collaboration measured non-zero cosine modulations
for positive pions~\cite{Osipenko:2008rv} produced by semi-inclusive DIS off the proton.
The \compass\ collaboration presented preliminary $\cstm$ and $\csm$ amplitudes in semi-inclusive DIS~\cite{Sbrizzai}
but has not yet published final results.
In Drell--Yan experiments non-zero 
azimuthal modulations have been measured~\cite{Falciano:1986wk,Guanziroli:1987rp,Conway:1989fs,Heinrich:1991zm,Zhu:2006gx,Zhu:2008sj}
that violate the Lam--Tung relation~\cite{Lam:1980uc}. 
Such a violation can be ascribed to the \BM\ distribution function, 
as pointed out in Ref.~\cite{Boer:1999si}. Sizable modulations 
have been extracted in pion-induced Drell--Yan reactions, 
where a valence quark and a valence antiquark annihilate.
When a sea parton is involved, as in proton-induced Drell--Yan processes, 
the measured modulations become smaller, suggesting a small \BM\ function for the sea.

This paper presents cosine modulations for positively and negatively charged
unidentified hadrons as well as for identified 
charged pions and kaons produced by DIS off hydrogen and deuterium targets.

%% file: analysis.tex
\section{The HERMES experiment}

The cosine modulations described in the previous section were
 extracted from measurements performed at the fixed-target 
\hermes\ experiment.
\hermes\ acquired data from  $1995$ to $2007$ with various polarized and 
unpolarized gaseous targets internal to
the \hera\ $27.6$~GeV electron/positron storage ring at \desy.
In this paper results are presented that were extracted using only the pure
hydrogen and deuterium targets, 
where the lepton beam scatters directly off neutrons
and protons (with only negligible nuclear effects in case of deuterium).
The spectrometer~\cite{Ackerstaff:1998av} was a forward-angle instrument 
consisting of two symmetric halves above and below the horizontal plane 
defined by the lepton-beam pipe.
Particles with polar angles within $\pm 170$ mrad in the horizontal direction and 
between $\pm(40$--$140)$ mrad vertically could be detected.
The collected data were processed with a
tracking code involving event-level fitting based on a 
Kalman-filter algorithm~\cite{Fruhwirth:1987fm},
which corrects the tracking parameters
for the effects from magnetic 
fields and accounts for all detector materials and known
mis-alignments.

Lepton-hadron separation with an efficiency better than $98\%$ was achieved
using the combination of several detectors:
a transition-radiation detector~\cite{Ackerstaff:1998av}, 
a dual-radiator ring-imaging Cherenkov (RICH) 
detector~\cite{Akopov:2000qi,Jackson:2005eu},
a lead and scintillator preshower detector, 
and a lead-glass calorimeter~\cite{Avakian:1998bz}.
Hadron identification was performed using
the RICH detector, taking into account the entire event topology simultaneously, 
rather than a single particle at a time.
This provides improved particle identification compared to earlier algorithms
(see appendix~\ref{sec:EVT} for further details).


\section{Data analysis}

\subsection{Data selection}\label{sec:Cuts}
The data used for this work were collected during the 2000-2007 periods,
with both lepton beam charges and unpolarized hydrogen and deuterium targets,
as summarized in table~\ref{tab:stat}.
In order to guarantee high-quality data, each event had to meet several criteria, 
such as good performance of the particle identification 
and tracking detectors. 
Each selected track satisfies geometric constraints to ensure that
it originated from the beam-target interaction region and also
remained well within the acceptance of the spectrometer.

\begin{table}[t]
\caption{Numbers of charged hadrons for each data set (in millions).}
\centering
\begin{ruledtabular}
\begin{tabular}{cccccc}
\multicolumn{6}{c}{\underline{Data set}}  \\
Year: & 2000 & 2005 & 2006 & 2007 &  \multirow{2}{*}{Total}\\
Beam: & e$^+$ &e$^-$ & e$^+$ & e$^+$ & \\
\\
\multicolumn{6}{c}{\underline{Hydrogen target}} \\
h$^+$ & 0.80 & -   & 1.97 & 2.11 & 4.88 \\
h$^-$ & 0.45 & -   & 1.12 & 1.20 & 2.77 \\
$\pi^+$ & 0.57 & - & 1.42 & 1.53 & 3.52 \\
$\pi^-$ & 0.40 & - & 0.99 & 1.07 & 2.46 \\
K$^+$ & 0.10 & -   & 0.24 & 0.26 & 0.60 \\
K$^-$ & 0.03 & -   & 0.08 & 0.09 & 0.20 \\
\\
\multicolumn{6}{c}{\underline{Deuterium target}} \\
h$^+$ & 1.02 & 0.52   & 0.48 & 0.55 & 2.57 \\
h$^-$ & 0.66 & 0.34   & 0.31 & 0.36 & 1.67 \\
$\pi^+$ & 0.72 & 0.38 & 0.35 & 0.40 & 1.85 \\
$\pi^-$ & 0.58 & 0.31 & 0.28 & 0.32 & 1.49 \\
K$^+$ & 0.12 & 0.06   & 0.06 & 0.07 & 0.31 \\
K$^-$ & 0.04 & 0.02   & 0.02 & 0.03 & 0.11 \\
\end{tabular}
\end{ruledtabular}
\label{tab:stat}
\end{table}

Events with at least one lepton and one hadron 
detected in coincidence are included in the semi-inclusive 
DIS event sample if they satisfy the following kinematic requirements.
The DIS region is defined here by the kinematic constraints
$Q^2>1$~GeV$^2$ and $W^2>10$~GeV$^2$, 
where $W^2$ is the squared invariant mass of the 
initial system of virtual photon and target nucleon.
As a consequence of these requirements and the limited angular acceptance of \hermes,
$x$ and $y$ are restricted to the ranges $0.023<x<0.6$ and $0.2<y$.
In addition the restriction $y<0.85$ is applied,
dictated by the energy threshold in the calorimeter to ensure 
high trigger efficiency.
In order to suppress hadrons not originating
from the struck quark (i.e., to suppress those from the 
\emph{target fragmentation region}),
the requirements $z>0.2$ and $x_F>0.2$ are applied. 
Here, $x_F=2p_{\scriptscriptstyle{z}}/\sqrt{s}$ 
is the Feynman scaling variable, where $p_{\scriptscriptstyle{z}}$
is the hadron momentum component parallel to the virtual photon, and 
$\sqrt{s}$ is the total energy in the $\gamma^* p$ center-of-mass system.

To ensure good identification of hadrons by the RICH,
identified pions are required to have momenta within $1$~GeV~$<P_{h}<15$~GeV
and kaons within $2$~GeV~$<P_{h}<15$~GeV.
RICH weights are assigned to each hadron.
These weights correspond to the probabilities that the hadron is a pion or a kaon,
as determined from the RICH hadron type hypothesis and 
the identification efficiency, computed from a \MC\ simulation 
of the RICH detector, which has been tuned to data.
No RICH identification or RICH weights are applied
for the data sample of unidentified hadrons.
To be consistent with the pion sample, 
the momentum restriction  $1$~GeV~$<P_{h}<15$~GeV
is also applied to unidentified hadrons.
In the calculation of all kinematic quantities that require particle masses, 
the pion mass was used for unidentified hadrons. 
This is as a good approximation as 
$70\%$ ($88\%$) of positive (negative) hadrons are pions (see table~\ref{tab:stat}).

The selected event sample is corrected for contamination
from leptons that do not originate from the scattered beam but rather come from lepton-pair production in detector material 
or meson Dalitz decay ($\pi^0/\eta \rightarrow \gamma e^+ e^-$).
These events amount to less than $1\%$ of the total number of events, 
and are typically concentrated toward the high-$y$ region,
where their contribution reaches $8\%$. 
These contaminating processes are charge symmetric. Therefore events passing DIS selection but with the wrong lepton charge 
constitute a control sample which is kinematically matched to
the background events wrongly included in the semi-inclusive DIS sample.
The correction is performed by assigning a negative weight to the events with 
lepton charge opposite to that of the beam.


\subsection{Extraction procedure}\label{sec:procedure}

Experimentally, the azimuthal modulations of the unpolarized cross section
can be accessed via the $\csNm$-moments ($n=1,2$)
\be\label{eq:moments}
\csN_\uu \,=\,\frac{\int_0^{2\pi} \csNm\,d{\sigma}_\uu\, {\mathrm d}\ph }{\int_0^{2\pi} d{\sigma}_\uu\,{\mathrm d}\ph},
\ee
where $d{\sigma}_\uu$ is defined in Eq.~\ref{eq:noncol_csec}.
The moments are related to the structure functions of interest 
via the $\ph$-independent part of the cross section
\begin{align}
\label{eq:MomStruc1}
 F_\uu^{\csm}  &= \frac{2\cs   }{\sqrt{2\epsilon(1+\epsilon)}} (F_\uut+ \epsilon F_\uul),\\
 F_\uu^{\cstm} &= \frac{2\cst  }{\epsilon}    (F_\uut+ \epsilon F_\uul).
\label{eq:MomStruc2}
\end{align}

Extracting the cosine modulations of the unpolarized cross section from data requires
disentangling them from a number of experimental sources of
azimuthal modulations.
At \hermes, due to the separation of the spectrometer into
two symmetric top-bottom halves, 
the azimuthal acceptance is non-uniform.
In addition, the observed kinematic conditions of each event may differ from
the Born conditions at the hard electromagnetic vertex due to both physical and 
experimental effects. Events may be reconstructed with altered kinematic conditions due to 
external bremsstrahlung and multiple scattering in the
detector material, or due to initial or final state radiation from the beam lepton (higher-order QED effects).
All of these effects lead to a miscalculation of the kinematic conditions, and
can induce false $\csNm$ modulations.

To correct the data for kinematic smearing and false cosine modulations,
a binned unfolding procedure was applied. 
A large  \pythia~\cite{PYTHIA6} \MC\ simulation 
(with approximately $20$ times more events than the experimental data) was generated,
which uses the \jetset~\cite{Sjostrand:1993yb} fragmentation model 
tuned to \hermes\ kinematic conditions~\cite{Airapetian:2010um}.
This simulation includes QED radiative effects calculated with \radgen~\cite{radgen},
and a complete \geant3~\cite{geant} simulation of the \hermes\ spectrometer.
All known instrumental and reconstruction effects are simulated,
including particle interactions with detector materials
and detector responses that account for known inefficiencies.
The simulated events then pass through the \hermes\ event-reconstruction algorithm,
mimicking any possible tracking bias or inefficiency present in the data.
The simulated semi-inclusive DIS sample provides information on both the Born-level
and the observed smeared kinematic conditions, and thus it can be used to
build a matrix  that describes the migration of events between kinematic bins.
This simulation is also used to define the events that smear into the
measurement from outside the accepted kinematic range, and thus represent
\emph{background events} in each kinematic bin.

A \emph{smearing matrix} is constructed by normalizing the
migration matrix to the Born cross section, taken from a Born-level 
\pythia\ production, in each bin.
As a result of this normalization, the smearing matrix is a relative quantity,
and reflects the fraction of events that are within the \hermes\
acceptance and their bin-by-bin migration. In the limit of
infinitely small bins, the smearing matrix is independent of the cross section model
used to build it. 
However, the kinematic distribution of the background events depends 
on the Born cross section model used to describe events generated outside the
\hermes\ acceptance.
The model dependence of the smearing matrix (due to finite sized bins) and
the background are discussed in section~\ref{sec:modeldep}.

The simulated samples are normalized relative to the data 
via the inclusive DIS cross section, as determined by the \lepto\ Monte Carlo generator~\cite{Ingelman:1996mq}. 
To correct the data for events smeared into the acceptance, 
the background events are subtracted from the normalized yields.
The smearing matrix is then used to unfold the
background-subtracted data, correcting for QED radiative effects, 
detector effects, acceptance and all the detector inefficiencies
included in the simulation.

The functional form
\be \label{eqn:ABC}
{\mathcal A} + {\mathcal B}\csm + {\mathcal C}\cstm
\ee
is fit to the azimuthal distribution of the unfolded yields 
to extract the cosine modulations $\cs_\uu=\mathcal{B}/2\mathcal{A}$ and $\cst_\uu=\mathcal{C}/2\mathcal{A}$.
As the  unfolding procedure is a linear operation, it can be 
combined with the linear operation of fitting.
This was done with linear regression, where a $\chi^2$ was formed:
\begin{align}
\chi^2= (\sigma^{\text{data}} - SX\beta)^T C^{-1} (\sigma^{\text{data}} - SX\beta).
\label{eq:FoldFit}
\end{align}
Here, $\sigma^{\text{data}}$ is the measured, background-subtracted yield,
$S$ is the smearing matrix, and $C$ is a covariance matrix
that includes the statistical uncertainties of data and background, and 
the statistical precision of the \MC\ used to construct the smearing matrix.
The product $X\beta$ gives the fit function of Eq.~\ref{eqn:ABC} 
representing the Born-level event yield,
with $\beta$ the vector of parameters (${\mathcal A}$, ${\mathcal
B}$, and ${\mathcal C}$) and $X$ the block diagonal \emph{design matrix}  that
includes a constant term (equal to $1$) and the values of 
$\cos\langle\ph\rangle$ and $\cos2 \langle\ph\rangle$  for each
$\ph$ bin.  See appendix~\ref{sec:FoldFit} for a detailed explanation of these matrices.

The running conditions at \hermes\ changed from year to year.
Thus the data of each year must be independently unfolded with the proper
\MC\ production, including the appropriate experimental configuration. 
To combine the results from the various data sets, the formalism of Eq.~\ref{eq:FoldFit} 
has been extended to a procedure that at the same time unfolds each data set
independently and fits the Born-level yields from all years
simultaneously. Technical details about the full procedure are
provided in appendix~\ref{sec:FoldFit}.

As the Born cross section depends on five kinematic variables
(see Eq.~\ref{eq:noncol_csec}), this 
procedure is carried out on a five dimensional grid of kinematic bins.
An analysis in fewer dimensions would implicitly integrate over variables 
and mix together physics and experimental effects. 
This mixing can obscure the true signal, as demonstrated by \MC\ tests.
A \MC\ simulation with an isotropic distribution in $\ph$ 
at the Born level was run through the detector simulation.
When unfolded in less than five dimensions, false modulations 
were extracted, which were of similar size as the physical 
moments seen in the data~\cite{GiordanoTr08}.

The binning used is reported in table~\ref{tab:Bins}. 
After the fit to the $\ph$-dependence, the final four-dimensional ($4D$) 
cosine modulations represent fully differential results.
Due to the unfolding procedure the results in the various kinematic 
bins are statistically correlated as well as the results for the 
$\cs$ and $\cst$ moments in each bin due to the fitting procedure. 
Therefore, the complete covariance matrix must be considered to avoid 
overestimating the statistical uncertainties in results projected on fewer dimensions.
\begin{table}[t]
\caption{Kinematic bin boundaries}
\centering
\begin{ruledtabular}
\begin{tabular}{rlllllll}
 x  : & 0.023 & 0.042 & 0.078 & 0.145 & 0.27 & 0.6 \\
 y  : & 0.2 & 0.3 & 0.45 & 0.6 & 0.7 & 0.85 \\
 z  : & 0.2 & 0.3 & 0.4 & 0.5 & 0.6 & 0.75 & 1.0 \\
 $\Pt$ [GeV]: & 0.05 & 0.2 & 0.35 & 0.5 & 0.7 & 1.0 & 1.3 \\
 $\ph$ : & \multicolumn{4}{l}{12 equidistant bins} \\
\end{tabular}
\end{ruledtabular}
\label{tab:Bins}
\end{table}

%% file: results.tex
\section{Systematic uncertainties} \label{sec:sys}     
This section discusses the systematic uncertainties
related to the imperfect treatment of
instrumental bias and inefficiencies by the smearing matrices. 
Systematic contributions related
to residual model dependence of the unfolding procedure due to 
finite bin sizes are also evaluated.

In contrast to the case of unidentified hadrons, the 
systematic uncertainties for 
identified pions and kaons include an additional 
contribution from the RICH identification.
In these cases, the \MC\ studies described below
include a full simulation of the RICH detector.

\subsection{Instrumental effects}\label{sec:sysinstr}
The geometric acceptance of the \hermes\ spectrometer produces
cosine modulations larger than the measured signals;
therefore a number of systematic checks have been performed
and are listed in this section.

The experimental apparatus experienced several major changes
over time. The lepton beam charge changed as \hera\ 
alternated between accelerating
electrons and positrons.
For the last two years of data taking, the target cell was shifted in
the beam-line direction closer to the forward spectrometer, 
and its length, initially of $40$ cm, was reduced by a factor of two.
Different magnetic fields were active in the target region 
in different years. The cosine modulations were extracted from
data collected  without any target magnetic field and with
longitudinal magnetic fields: a solenoid of $0.3$
Tm strength, employed for a longitudinally polarized target in 2000,
and the $1.0$ Tm solenoid of the \hermes\ recoil detector
installed after 2005.
In addition, during shut-down periods some detectors
were moved in and out, and relative positions between detectors changed. 
All these altered conditions induce changes in the geometric
acceptance; therefore each data taking period requires a dedicated
simulation to properly correct for the acceptance.

Despite those significant changes in running conditions,
the cosine modulations extracted separately for each year
are found to be mostly consistent.
Small systematic shifts between years are observed,
which can be ascribed to effects not included in the simulations, 
and thus in the correction. 
These effects include residual detector misalignment not accounted for
in the tracking algorithm, which are expected to change from one 
data-taking period to another.

The time stability of the apparatus response was 
checked by measuring the azimuthal modulations generated by the 
\hermes\ acceptance in short time intervals within the same data 
taking period. 
The test indicated that the azimuthal modulations of the acceptance are stable
in time. 

The tiny instabilities ($<1\%$ of the observed amplitudes) are highly 
dependent on the year under study, and, as the \MC\ does not simulate or
correct for any of these instabilities, they can partially explain 
the small differences between cosine modulations extracted from 
different years.

To take these differences into account,
the signed difference between moments extracted from each year
and moments extracted from a simultaneous fit of the remaining periods was 
evaluated at the $4D$-level. The modulus of the weighted average of these 
differences was added to the systematic uncertainty, and
represents the largest contribution to the systematic uncertainty
($\sim$$70\%$ of the total systematic uncertainty).

The extracted moments were checked for
a sensitivity to a possible beam 
misalignment or slope with respect to the spectrometer axis, 
and misplacement of the spectrometer dipole magnet.
No significant effects have been found. The net beam polarization
was found to be negligible. 
Additional instrumental sources that could generate false azimuthal 
modulations have been tested by measuring sine modulations and cosine 
modulations higher than $\cstm$, which are not present 
in the unpolarized semi-inclusive DIS cross section in the single-photon
exchange approximation. No significant signals were found.

The final moments discussed in section~\ref{sec:fully} and~\ref{sec:fixedRange}
have not been corrected for possible binning effects 
or for RICH inefficiencies or cross-contaminations that were not 
accounted for (for identified pion and kaon samples only),
for example due to $\ph$-dependence not accounted for in the RICH weights. 
The influence of these effects on the final moments was estimated by a \MC\ simulation. 
For each particle type under study, a $4D$ model 
of the measured cosine modulations was extracted from the fully differential final moments
by means of a $4D$ parameterization (details in appendix~\ref{sec:Param}).
Through an accept/reject procedure, those models were used to 
alter, at the Born-level, the underlying distribution in an 
originally azimuthally uniform \pythia\ production that includes the full
spectrometer simulation and QED radiative effects. For the 
identified hadron cases, 
both pion and kaon models 
were implemented, to account for cross-contaminations between modulations.
The protons constitute the remaining significant
part of the hadron sample, but a model was not extracted for them.
The sensitivity of the test to proton modulations was checked by implementing  
a model for protons that was either $\ph$-independent or with modulations identical
to that of pions. The proton model
input to the \MC\ is found to have very little impact on the 
test results.

This simulation, modeled to reproduce the measured cosine azimuthal 
modulations, was used as a surrogate for the data in the entire analysis
procedure, and cosine modulations for pions and kaons and unidentified hadrons were extracted.
The extracted moments were found to agree with the input models,
verifying that the unfolding algorithm is able to extract the implemented modulations
after correcting for all instrumental and QED radiative effects included in the simulation.
The small discrepancies between the extracted moments and the input model
provide an estimate of systematic uncertainty 
due to the unfolding procedure, 
binning effects and RICH weights (in the case of the identified pion and kaon
samples).

\subsection{Model dependence}\label{sec:modeldep}
The unfolding procedure described in section~\ref{sec:procedure}
can be affected by two different sources of model dependence.
The unfolding correction is based mainly on two objects:
a smearing matrix, describing the migration of events
between bins, and a background estimation, describing the events that are 
smeared into the kinematic bins from outside the geometric/kinematic acceptance.

In a fully differential analysis and in the limit of infinitely 
narrow bins, the smearing matrix is independent of the models underlying the Monte Carlo
event generator used to produce it.
Residual model dependence due to finite bin sizes 
was tested by comparing data azimuthal moments extracted using 
smearing matrices computed with different models for the azimuthal
dependent part of the cross section:
the standard, $\ph$-independent, \pythia\ cross section, 
and the altered \pythia\ cross section that includes
the $4D$ cosine model extracted from data, 
as described in section~\ref{sec:sysinstr}.
As expected, no significant differences
were observed in the extracted moments.
 
To test the model dependence of the background a similar 
procedure was used, and cosine modulations extracted 
with different models for background evaluation were compared.
Again the two models used were 
the standard, $\ph$-independent \pythia\ cross section, 
and the \pythia\ cross section modified to include
the $4D$ cosine model, 
which was extrapolated into the unmeasured region not covered by the detector acceptance.
The moments from data
 were found to be weakly sensitive to the 
azimuthal dependence of the background.
The differences between the moments extracted 
with the two models were used to estimate the systematic uncertainties
from the model dependence and were combined with the other 
systematic uncertainties in quadrature.\\

\subsection{Calculation of systematic uncertainties}\label{sec:calcsys}
The systematic uncertainties described above are subject to statistical fluctuations
due to the finite statistical precision of the \MC\ simulations
used to calculate them.
To average out these statistical fluctuations, each systematic contribution
was smoothed by fitting it to a $4D$ linear function.
Higher order $4D$ polynomials were tested, and provided 
final systematic uncertainties of similar size.
The final systematic uncertainty was calculated by adding each smoothed 
contribution in quadrature.


\section{Fully differential results}\label{sec:fully}
The $4D$ analysis described in section~\ref{sec:procedure} 
provides access to the full kinematic dependences.
The final moments, in four dimensions, 
for positive and negative unidentified hadrons, pions, and kaons 
produced from hydrogen and deuterium targets,
are available online~\cite{Database}.
These fully differential moments represent the 
complete set of results of this analysis and can be used to test theoretical models.

The moments are accompanied by the covariance matrix describing 
the statistical correlations, the total systematic uncertainties for each bin 
(all contributions are added in quadrature), and the average
$\la x\ra$, $\la y\ra$, $\la z\ra$, $\la \Pt\ra$, 
and $\la Q^2\ra$ values for each bin.
It is not possible to make a measurement in every ($x$, $y$, $z$, $P_{h\perp}$) bin due to
(a) kinematic constraints that exclude some portions of the $4D$ space, 
(b) the not-uniform distribution of the underlying cross section across the rectangular kinematic binning,
(c) the detector acceptance, and
(d) the limited statistical precision of the data.
Bins that do not contain enough events to make a measurement
(and typically also have statistical uncertainties larger than unity)
are denoted in the database with all moments and average kinematics set to zero.  
In the covariance matrix, elements corresponding to
such bins have diagonal element values of one and non-diagonal
element values of zero.

A visual representation of the bins where a measurement is not possible can be found online~\cite{cherrypicker}
for the statistically poorest data set for each particle type (pions, kaons, and all hadrons).
This tool also allows the user to integrate the moments in an arbitrary kinematic range (following the procedure described in the next section).

\section{Results for fixed kinematic ranges}\label{sec:fixedRange}
\begin{table}[t]
\caption{Kinematic ranges of integration.}
\centering
\begin{ruledtabular}
\begin{tabular}{c c c c}
\multicolumn{4}{c}{}\\
\multicolumn{4}{c}{Kinematic Range A} \\
$x$ & $y$ & $z$ & $\Pt$ [GeV] \\
0.023 - 0.27  &  0.3 - 0.85  &  0.2 - 0.75  &  0.05 - 1.0 \\
\multicolumn{4}{c}{}\\
\multicolumn{4}{c}{Kinematic Range B} \\
$x$ & $y$ & $z$ & $\Pt$ [GeV] \\
0.042 - 0.27  &  0.3 - 0.7  &  0.2 - 0.6  &  0.2 - 0.7 \\
\end{tabular}
\label{tab:KineRange}
\end{ruledtabular}
\end{table}
The fully differential moments provide the maximum information 
from this measurement.
In order to gain a qualitative 
picture of the behavior of the moments, a 
projection to one dimension ($1D$) was performed by
a weighted integration of the moments over three variables,
highlighting the dependence of the moments on the fourth variable. 
In order to achieve an integral over the selected kinematic ranges,
the moment in each kinematic bin is folded with the $\ph$-integrated unpolarized 
	semi-inclusive DIS cross section in this bin, normalized 
	to the same value integrated over the whole kinematic range of the 
	projection. 
The necessary input cross section for this integration was extracted directly from \hermes\ data~\cite{dc19}.

\begin{figure*}[htp]
\begin{center}
\includegraphics[width=0.8\textwidth]{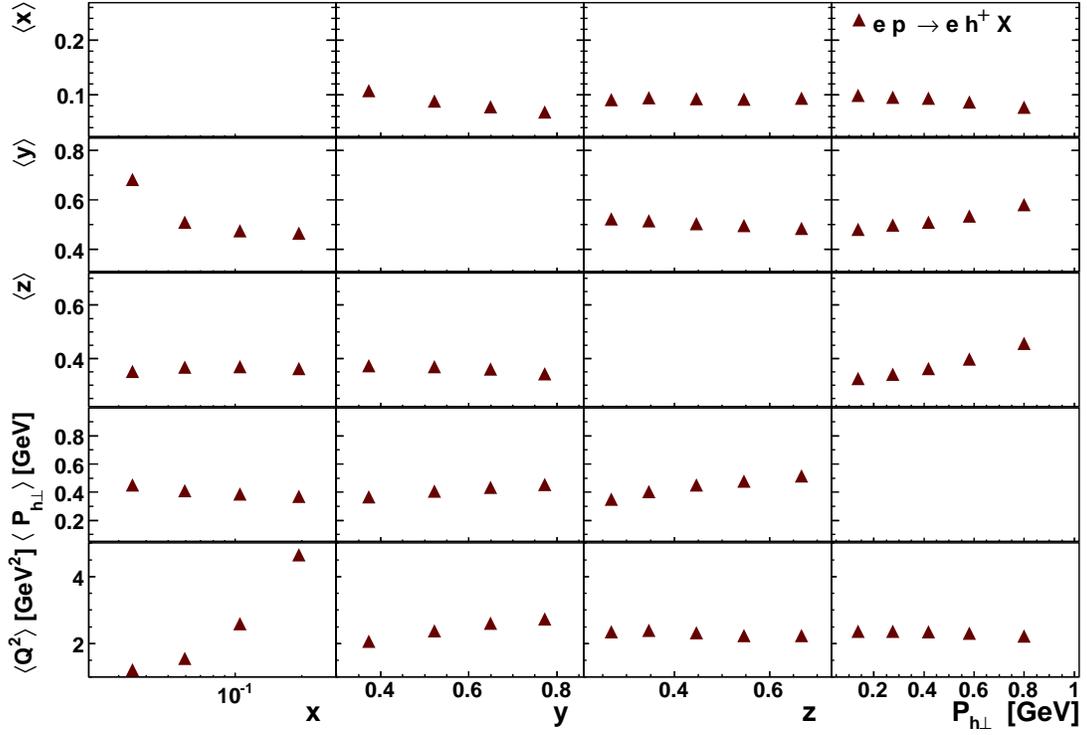}
\caption{Average kinematics for the integration range
	 \emph{A} of table~\ref{tab:KineRange}, as extracted from a $4\pi$ Monte Carlo 
        (shown here for positive hadrons 
	on hydrogen; other cases exhibit only minor deviations).}
\label{fig:AvgA}
\end{center}
\end{figure*}

\begin{figure*}[h!btp]
\begin{center}
\includegraphics[width=0.8\textwidth]{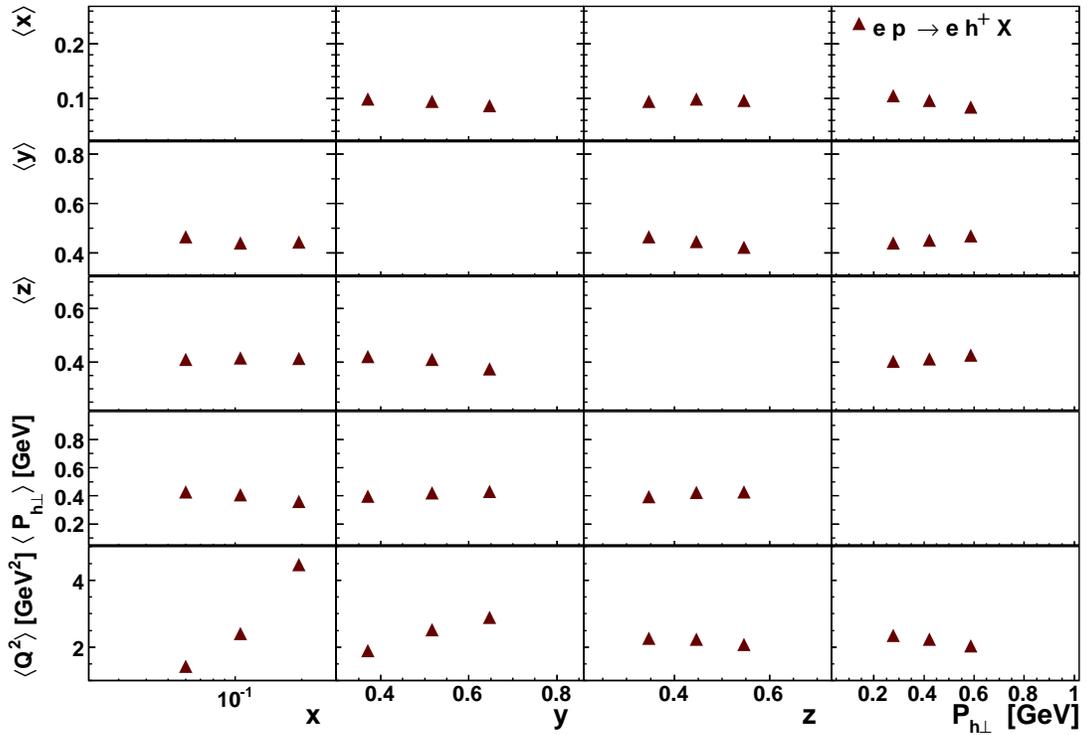}
\caption{As in Fig.~\ref{fig:AvgA}, but for the kinematic range \emph{B} of
  table~\ref{tab:KineRange}.}
\label{fig:AvgB}
\end{center}
\end{figure*}

As anticipated in section~\ref{sec:fully}, it is not possible to make a measurement in every bin.
Therefore, restricted
kinematic ranges of integration were chosen 
to minimize the number of bins in the integration where
a measurement is not possible.
In addition, the $z$-bin from $0.75$ to $1.0$ is excluded as
a large fraction of events in this kinematics range contain
decay products of exclusively produced hadrons, 
for which standard factorization might be broken.
Hadron and pion results are integrated over the kinematic range \emph{A}, 
given in table~\ref{tab:KineRange}. This table also lists
the reduced kinematic range {\emph B} used for kaons,
which have comparatively lower statistical precision.
The average kinematic values for each integrated 
bin for the ranges {\emph A} and {\emph B}
are shown in Figs.~\ref{fig:AvgA} and~\ref{fig:AvgB},
respectively. 
The strong correlation between $x$ and $Q^2$ (and thus between $x$ and $y$)
shown in the bottom left subplot 
is due to the \hermes\ acceptance.  
The other kinematic variables are weakly correlated.

In the integration, bins with large statistical uncertainties
(larger than unity) are not included as they do not
provide meaningful information to the integral 
and inflate its uncertainty. The effect of excluding these 
bins is estimated and included in the 
systematic uncertainty of the projected results, as described at the end of this section.
In figures that compare results from various hadron types or
charges, the ranges of integration have been chosen 
so that only the bins that provide a measurement in the statistically poorest data set
are included.
This assures that the data sets have identical integration regions 
and thus allows for the results to be compared in a consistent way.

The integrated azimuthal modulations were found to be weakly sensitive 
to the semi-inclusive DIS cross section used for the integration.
This sensitivity was assessed by using
the cross section extracted from \hermes\ data as well as the cross sections implemented in
two \MC\ simulations tuned to reproduce the \hermes\ measured yields: \pythia~\cite{PYTHIA6}
and {\sc gmc\_{trans}}.
The \hermes\ {\sc gmc\_{trans}}
generator uses the \cteq6 distribution functions~\cite{Pumplin:2002vw},
and the DSS fragmentation functions~\cite{deFlorian:2007aj,deFlorian:2007hc}
with the $\Pt$-dependence based on a Gaussian ansatz.
In particular, the transverse momenta $p_T$ have a non-constant
$z$-dependence as observed from a fit to \hermes\ data
~\cite{Schnell:2007}, while for \pythia\ this $z$-dependence is flat.
The sensitivity to semi-inclusive DIS cross section used for the integration
has been added to the systematic uncertainty.

The systematic uncertainties for the results projected in $1D$ 
are composed of the uncertainties discussed in section~\ref{sec:sys}, the sensitivity to the cross section used in the integration described in the previous paragraph, plus the additional uncertainty 
added by excluding some bins from the projection.
The \MC\ production modified to reproduce the measured azimuthal distribution
(described in section~\ref{sec:sysinstr}) was used
to evaluate the effect of the bins excluded from the integration.
The difference between including or excluding these bins in the integration of the simulated moments
was added to the other systematic contributions.
Each systematic contribution was independently projected onto the single variable
before the smoothing described in section~\ref{sec:calcsys}.
After the projection, the systematic contributions were smoothed with a $1D$ linear 
fit and then added in quadrature.

\subsection{Results for charged pions}\label{sec:pions}

The cosine modulations for charged pions, projected in the kinematic range 
{\emph A} (table~\ref{tab:KineRange}),
are presented in this section. 
All pion samples are projected only including bins that provide a measurement in every data sample,
which restricts the integration to those bins with a measurement
in the statistically poorest data sample,
i.e., the sample for negative pions produced from the deuterium target.

\subsubsection{\emph{Pion} $\cstm$ \emph{amplitudes}}

\begin{figure*}[htp]
\begin{center}
\includegraphics[width=0.95\textwidth]{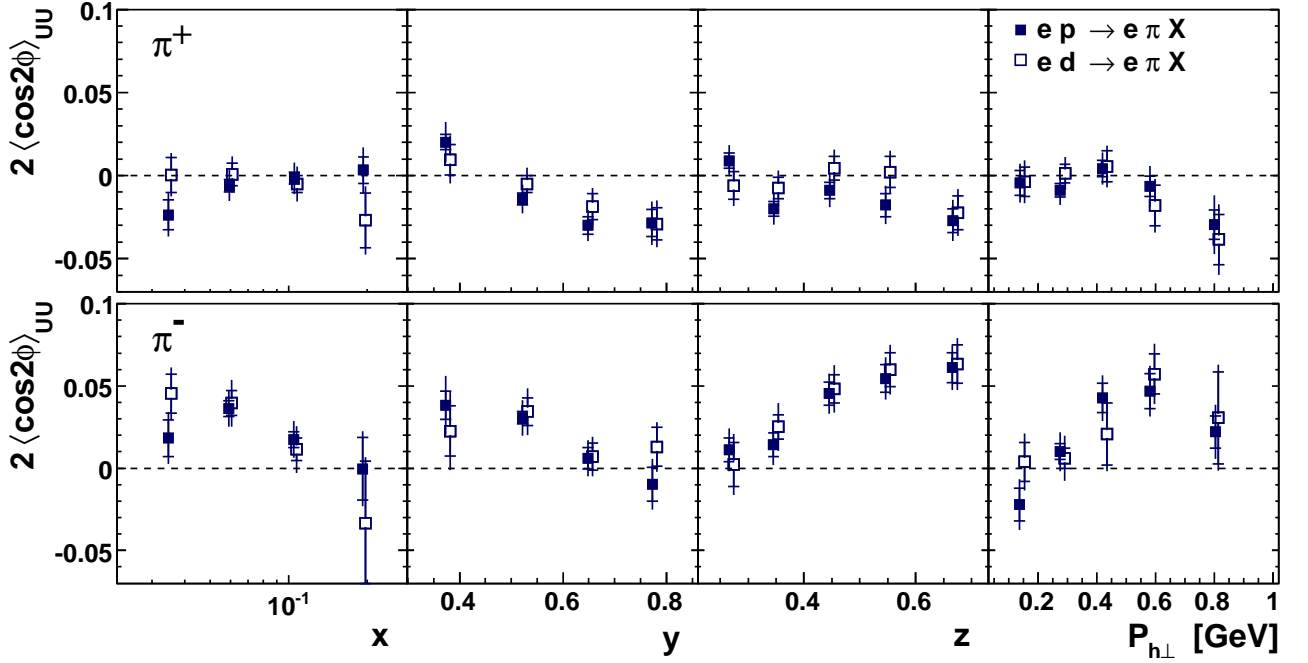}
\caption{$\cstm$ amplitudes for positive (upper panels) and negative
  (lower  panels) pions integrated over the kinematic range \emph{A} of table~\ref{tab:KineRange}. 
Closed and open squares are for amplitudes extracted from
  hydrogen and deuterium targets, respectively. The inner bar
  represents the statistical uncertainty; the outer bar is the total uncertainty, 
  evaluated as the sum in quadrature of statistical and systematic uncertainties.
Points have been slightly shifted horizontally for visibility.}
\label{fig:Pionscos2}
\end{center}
\end{figure*}

Figure~\ref{fig:Pionscos2} shows the $\cstm$ amplitudes $2\cst_\uu$ for pions
extracted from hydrogen and deuterium data, projected versus $x$, $y$, $z$, and $\Pt$.
Different magnitudes and opposite signs of the amplitudes are observed for oppositely charged pions. 
In particular, positive $\cstm$ amplitudes are extracted for negative
pions, while for positive pions the moments are compatible with zero, but tend to be negative in some kinematic regions.
The amplitudes for positive and negative pions also 
exhibit different kinematic dependences. 
This is particularly evident in their dependence on $z$: in the integrated kinematic region 
presented here, the magnitudes for positive pions have no clear kinematic dependence, while they rise with $z$ for negative pions.
The amplitudes increase in magnitude with $\Pt$ for both $\pi^+$ and $\pi^-$, but with opposite signs.

Up to a kinematic suppression of $(1/Q)$, 
the $\cstm$ amplitudes only contains a single, unsuppressed term, the \BMC\ effect, 
i.e., the convolution of the \BM\ distribution function  $h_1^{\perp}(x, p_T ^2)$ and the 
Collins fragmentation function $H_1^{\perp}(z,k_T ^2)$ discussed in section~\ref{sec:intro}. 
For a hydrogen target, scattering off $up$ quarks is expected to
dominate the reaction ($u$-dominance), both because the proton consists of
more $up$ quarks than $down$ quarks,
and because the elementary lepton-quark cross section is 
proportional to the squared quark electric charge ($e_q^2$),
which gives an additional factor of $4$ for
$up$ quarks compared to $down$ quarks.
The Collins function was recently found to have a similar magnitude
but opposite sign for fragmentation of $up$ quarks into positive 
({\it favored fragmentation}) and negative pions 
({\it disfavored fragmentation})
~\cite{Aut2005,Anselmino:2007fs,Airapetian:2010ds,Alexakhin:2005iw,Abe:2005zx}.
This would result in  different signs for pions of opposite charge,
which is in agreement with the data.

The similarity between hydrogen and deuterium results seems to indicate that the \BM\ distribution function 
has the same sign for $up$ and $down$ quarks, as shown in Ref.~\cite{RebeccaThesis} and in Ref.~\cite{Barone:2009hw}, 
and anticipated in Refs.~\cite{Burkardt:2007xm,Burkardt:2005hp}.
Although they are similar, for positive pions
the deuterium results seem to be systematically closer to zero with respect to the hydrogen results;
this might be due to a different magnitude of the \BM\ function for up and down quarks.

Model calculations \cite{Gamberg:2003ey,Gamberg:2007wm,Barone:2008tn,Zhang:2008ez} of 
the contribution of the \BMC\ effect to the $\cst_\uu$ moment are 
in qualitative agreement with the moments reported here. 
In particular, the opposite sign for oppositely charged pions
seems to be a signature of the Collins effect.

Equation~\ref{eq:Twist2} only includes terms up to a suppression $(1/Q)$, 
but at a suppression of $(1/Q)^2$ there is at least one additional term that includes the \C\ effect (see Eq.~\ref{eq:Twist4}).
The restricted $Q^2$ range of the \hermes\ data does not allow for a
conclusive study that disentangles the leading term from the suppressed terms.
Nonetheless, an attempt to describe preliminary \hermes\ results in a more complete way has been done
in Ref.~\cite{Barone:2008tn}, where the authors evaluated this suppressed 
\C\ contribution to the $\cstm$ amplitude, assuming a flavor-blind \C\ term, i.e., 
a flavor-independent $\la p_T^2 \ra$.
The comparison of this calculation to data indicates that, in the \hermes\ kinematic regime, 
the \C\ term is smaller than expected or is counteracted by additional terms that have been neglected.
In the same paper, a possible \C\ flavor-dependence was also estimated
by varying the $\la p_T^2 \ra$ for $down$ quarks while maintaining a fixed $\la p_T^2 \ra$ for $up$ quarks; 
no significant changes were observed in the calculated \C\ term.
However, this test was performed on a hydrogen target,
and not a deuterium target where the results might be more sensitive to the $down$ quarks.

In Ref.~\cite{Barone:2009hw}, the authors attempted to simultaneously describe  
preliminary unidentified hadron $\cstm$ amplitude extracted at 
\hermes~\cite{Giordano:2009hi} and \compass~\cite{Sbrizzai}.
The \BMC\ effect is described using the Collins \ff\ 
from Ref.~\cite{Anselmino:2007fs}
while for the \BM\ function, the same functional 
form that was used for the Sivers function~\cite{Anselmino:2005ea} was applied.
In the calculation the \C\ effect is also included, 
which is sensitive to the quark average transverse momenta.
The previously reported average momentum of $\langle p_T^2\rangle=0.25$~GeV$^2$~\cite{Anselmino:2005nn} describes the \compass\ data well.
In contrast, the \hermes\ data is better described by the lower value of $\langle p_T^2\rangle=0.18$~GeV$^2$,
leading to a smaller \C\ effect at \hermes.
This is in accordance to the broadening of the $p_T$ distribution when considering $Q^2$ evolution,
as observed in Ref.~\cite{Aybat:2011zv}. 

\subsubsection{\emph{Pion} $\csm$ \emph{amplitudes}}
\begin{figure*}[htp]
\begin{center}
\includegraphics[width=0.95\textwidth]{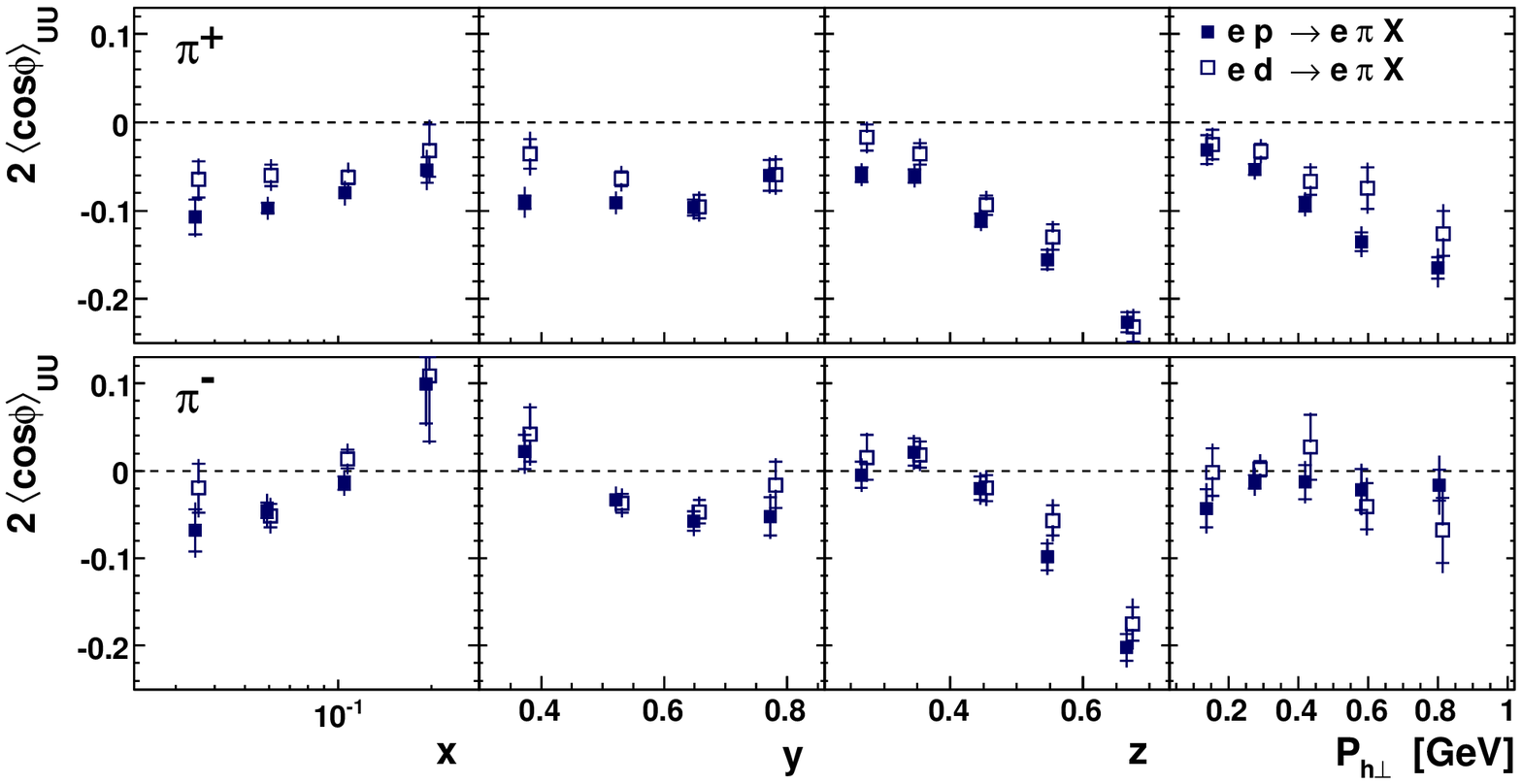}
\caption{As in Fig.~\ref{fig:Pionscos2}, but for the $\csm$ amplitudes.}
\label{fig:Pionscos}
\end{center}
\end{figure*}

The $\csm$ amplitudes come suppressed as $1/Q$ in the hadron cross section,
and, in contrast to the $\cstm$ amplitudes, several terms contribute at same level of suppression
(Eq.~\ref{eq:Twist3}).
Results for the $\csm$ amplitudes $2\cs_\uu$ extracted for 
pions from hydrogen and deuterium data are shown in Fig.~\ref{fig:Pionscos}. 
Results extracted from hydrogen and deuterium are similar,
but deuterium results for positive pions are 
smaller than hydrogen results. This could be related to 
flavor dependence of the contributions involved in the amplitudes.
The $\csm$ amplitudes are found to be negative for both positively and negatively charged pions,
but for positive pions they are in general larger in magnitude.
For both positive and negative pions, the magnitudes increase with the 
pion energy fraction $z$. 

The $z$ dependence of the amplitudes can be interpreted in terms of the \C\ effect.
Indeed, \C\ anticipated a rise of amplitudes with $z$ due to the reduced dilution 
by the random transverse momentum that the pions acquire during 
fragmentation~\cite{Cahn:1978se,Cahn:1989yf}. 
At high $z$ the amplitudes for oppositely charged pions are very similar 
and reach their largest magnitude (up to $-0.2$).
Different behaviors are observed for oppositely charged pions
versus $\Pt$.
The magnitude of the amplitudes for positive pions increases with $\Pt$, supporting the 
\C\ expectations of a signal proportional to transverse momentum.
But, this trend is not observed for negative pions, suggesting that 
contributions to Eq.~\ref{eq:Twist3} other than \C\ possibly counterbalance the increase 
with $\Pt$.
The \C\ term is expected to be weakly sensitive to flavor, 
as discussed in the previous section. 
As in the case of the $\cstm$ amplitudes, the difference between oppositely charged pions
can be generated by flavor dependent contributions, like, e.g., the \BMC\ effect.

In contrast to the $\cstm$ amplitudes, no model can qualitatively describe the
measured amplitudes for $\csm$. To date, only one prediction has been published 
for the \hermes\ $\csm$ amplitude~\cite{Anselmino:2006rv}, which includes the \C\ term only. 
The amplitudes predicted are larger than the measurements, 
suggesting that the \C\ contribution at \hermes\ is smaller than expected.
As in the case of $\cst_\uu$, at least part of the discrepancy can be related to a 
$\la p_T^2\ra$ that, in \hermes\ kinematic conditions, is smaller than $0.25$~GeV$^2$.
Moreover, the modeled \C\ term cannot describe the observed 
difference between $\pi^+$ and $\pi^-$, as it was assumed to be flavor-blind.
This implies that for a qualitative description of the measured 
$\csm$ amplitudes, a more complex \C\ contribution, 
or additional flavor dependent contributions, like, e.g., the \BMC\ effects, must be taken into account.
Furthermore, in addition to the Cahn and the Boer-Mulders terms, 
the structure function $F_\uu^{\csm}$ includes four terms related to quark-gluon-quark correlators
which have not been taken into account in this interpretation of the data, 
as little is known about the underlying physics.

\subsection{Results for charged kaons}

This section presents the cosine modulations extracted for charged kaons,
projected in the reduced kinematic range {\emph B} 
of table~\ref{tab:KineRange}.
All kaon samples are projected using bins that provide 
a measurement in the negative kaon sample produced from the deuterium target,
which is the statistically poorest kaon data sample.
No model calculation for kaons is available to date.

\subsubsection{\emph{Kaon} $\cstm$ \emph{amplitudes}}

\begin{figure*}[htp]
\begin{center}
\includegraphics[width=0.95\textwidth]{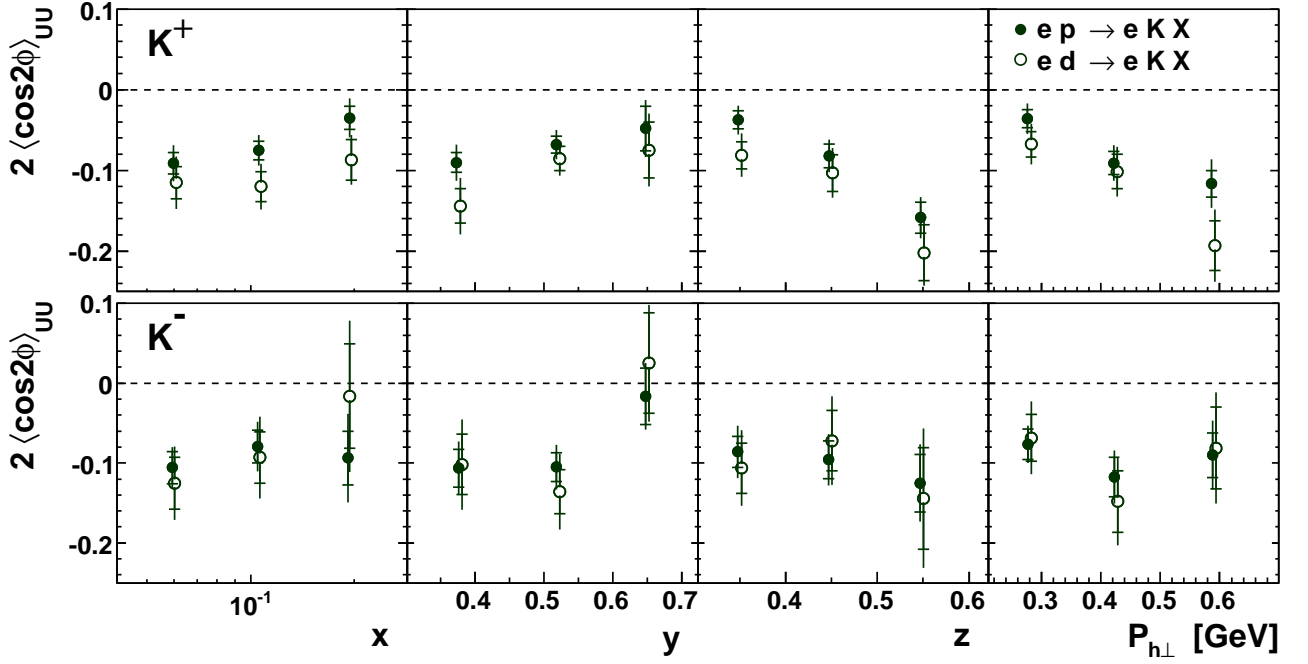}
\caption{As in Fig.~\ref{fig:Pionscos2}, but for charged kaon amplitudes integrated over the kinematic range \emph{B} of table~\ref{tab:KineRange}.}
 \label{fig:Kaonscos2}
\end{center}
\end{figure*}

The $\cstm$ amplitudes extracted for charged kaons are shown in 
Fig.~\ref{fig:Kaonscos2} for the hydrogen and deuterium targets. 
The amplitudes are large in magnitude (up to $-0.1$),
and have the same negative 
sign for both positive and negative kaons, in contrast to the trends observed for pions.
This may be interpreted by considering the kaon's quark content:
the valence quark content of $K^+$ mesons is $u \bar{s}$, and therefore $K^+$
production is expected to receive a large contribution from 
lepton scattering off $up$ quarks ($u$-dominance).
A favored Collins fragmentation function is expected to
be involved in this case, as in the case of $\pi^+$.
In the framework of the Artru model~\cite{Artru:1997bh}, 
all favored Collins functions describing fragmentation
into spin-zero mesons have the same sign.
Therefore, the \BMC\ effect for positive pions and kaons is
expected to have the same sign, as observed in measurements.
Nothing is known about the Collins fragmentation function 
into kaons. 
A significant contribution from sea quarks cannot be excluded.
For example, $strange$ quarks may contribute at $x<0.1$~\cite{Airapetian2008qf},
as suggested by unpolarized fragmentation, where the 
$strange$ quark fragmentation function into $K^+$ appears to be significantly larger than the fragmentation function for
$up$ quarks into $K^+$~\cite{deFlorian:2007aj}.
A substantial difference between the $strange$ and the $up$ and $down$ 
Collins fragmentation functions would play an important role in the observed moments.

For negative kaons the situation is even more complicated, as its valence quark
content ($s\bar{u}$) does not include any quarks in common with the valence structure
of the target.
Therefore, even larger contributions can be expected to originate
from the sea and from disfavored $up$ quark fragmentation.

Similar kaon amplitudes are extracted from hydrogen
and deuterium targets. This may reflect similar contributions from $u$ and $d$ quarks, as well
as a potentially substantial contribution from $strange$ quark fragmentation,
which is expected to be the same for neutron and proton targets.
Contrary to pions, the positive kaons show 
hydrogen results closer to zero, which might reflect a different magnitude for the \BM\ function
of different quark types, or the increased role of disfavored $up$ quark fragmentation for proton targets.

\subsubsection{\emph{Kaon} $\csm$ \emph{amplitudes}}

%

\begin{figure*}[htp]
\begin{center}
\includegraphics[width=0.95\textwidth]{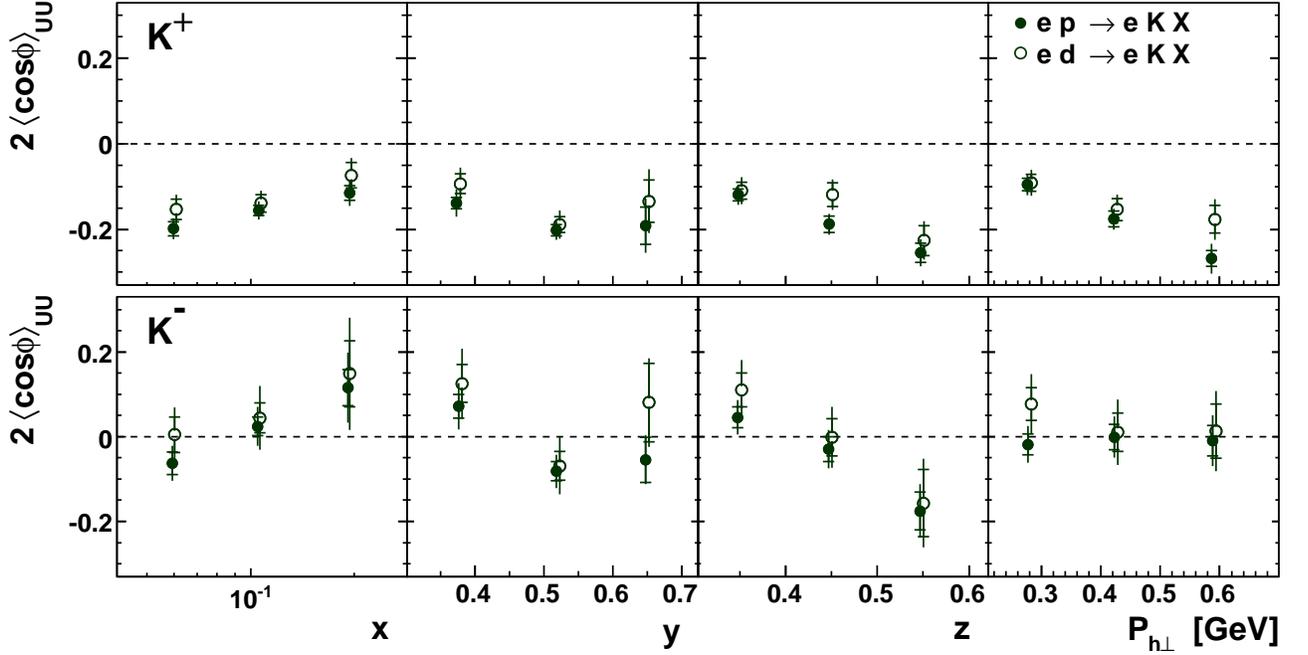}
\caption{As in Fig.~\ref{fig:Pionscos}, but for charged kaon amplitudes integrated over the kinematic range \emph{B} of table~\ref{tab:KineRange}.}
\label{fig:Kaonscos}
\end{center}
\end{figure*}

The $\csm$ amplitudes for kaons are shown in Fig.~\ref{fig:Kaonscos} 
for hydrogen and deuterium targets.
Large negative (up to $-0.2$) amplitudes are extracted for positive kaons,
slightly rising with $z$ and $P_{h\perp}$. 
The amplitudes are even larger in magnitude than those for positive pions, 
which suggests a large contribution 
from the \BMC\ effect,
which was found to be large for $K^+$ in the previous section.
Negative kaons instead show results compatible with zero.
The similarity between the $\cstm$ amplitudes for positive and negative kaons may mean that 
the \BMC\ effect is relatively insensitive to kaon charge.
Thus, the significant difference in the $\csm$ amplitudes for positive and negative kaons points to either 
a flavor dependence of the \C\ contribution (e.g. from $strange$ quarks) or
a significant contribution from the interaction dependent terms that have been otherwise neglected in this discussion.
Similar results are extracted for scattering off hydrogen and deuterium.

\subsection{Results for unidentified charged hadrons}\label{sec:hadrons}

In this section the cosine modulations extracted
for unidentified hadrons and projected in kinematic range {\emph A} 
(table~\ref{tab:KineRange}) are presented.
As for identified charged hadrons, individual kinematic bins are included in the integration  
only if they provide a measurement in the statistically poorest unidentified hadron
data sample, i.e., negative hadrons produced from a deuterium target.

As the majority of the unidentified hadrons consists of pions ($\gtrsim$$70-88\%$, depending on the hadron's charge),
the amplitudes of unidentified hadrons are very similar to those of pions,
and most of the arguments from the discussion of the pion results
also apply here. 
However, as no hadron identification was required, 
the systematic uncertainty for the unidentified hadron sample does not include a contribution from the RICH identification.
The remaining hadrons are in large part kaons ($\sim$$10\%$), 
and protons ($\sim$$10\%$).
As no theoretical model has evaluated the cosine modulations for kaons and protons,
no predictions exist for the unidentified hadron sample.

\subsubsection{\emph{Hadron} $\cstm$ \emph{amplitudes}}

\begin{figure*}[htp]
\begin{center}
\includegraphics[width=0.95\textwidth]{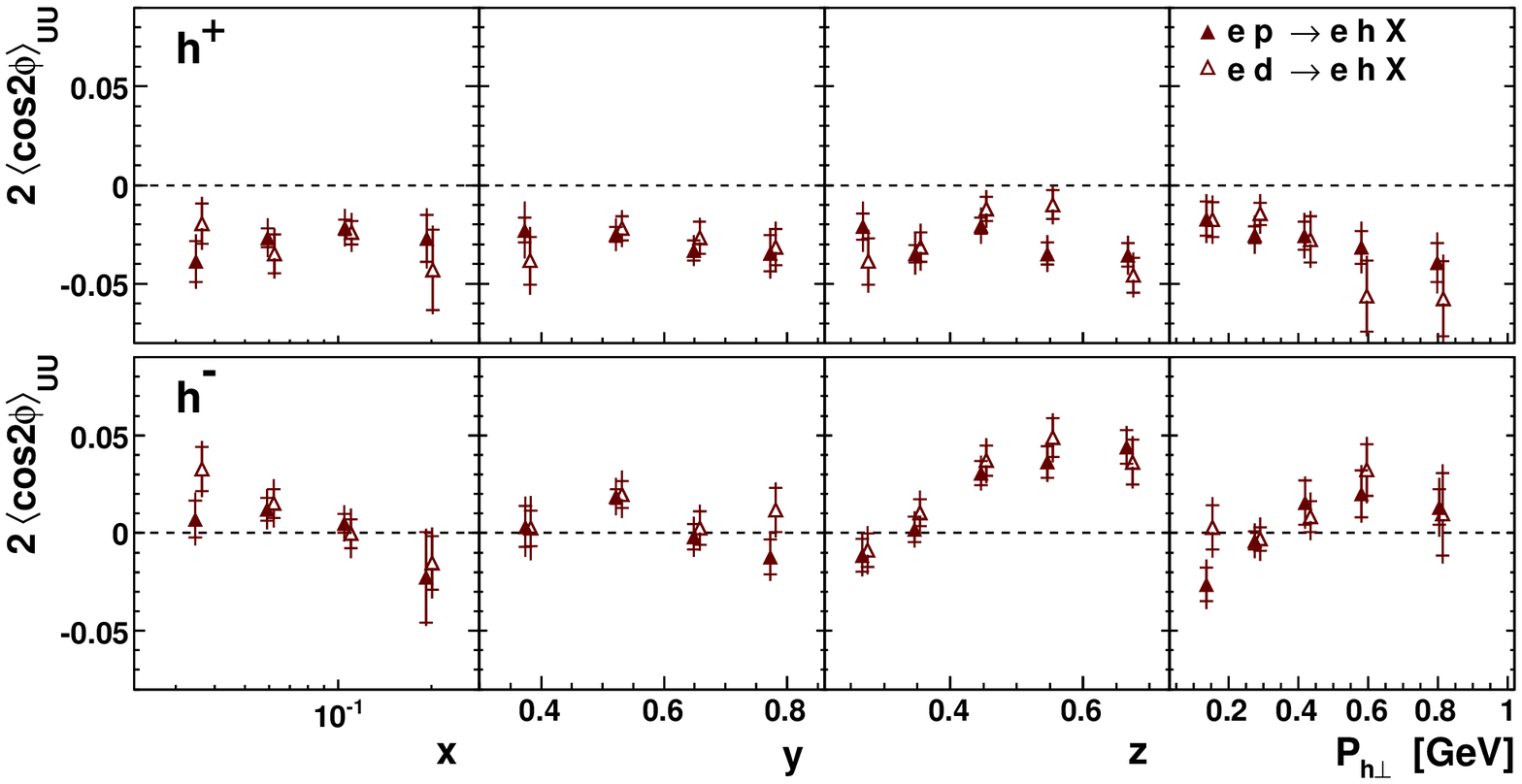}
\caption{As in Fig.~\ref{fig:Pionscos2}, but for unidentified charged hadrons.}
\label{fig:Hadronscos2}
\end{center}
\end{figure*}

Figure~\ref{fig:Hadronscos2} shows the $\cstm$ amplitudes
of unidentified hadrons extracted from hydrogen and deuterium data.
Different amplitudes are extracted for oppositely charged hadrons; 
in particular, they are of opposite sign, as in the case of pions.
Similar amplitudes are observed for hadrons produced from hydrogen and
deuterium targets.

\subsubsection{\emph{Hadron} $\csm$ \emph{amplitudes}}

\begin{figure*}[htp]
\begin{center}
\includegraphics[width=0.95\textwidth]{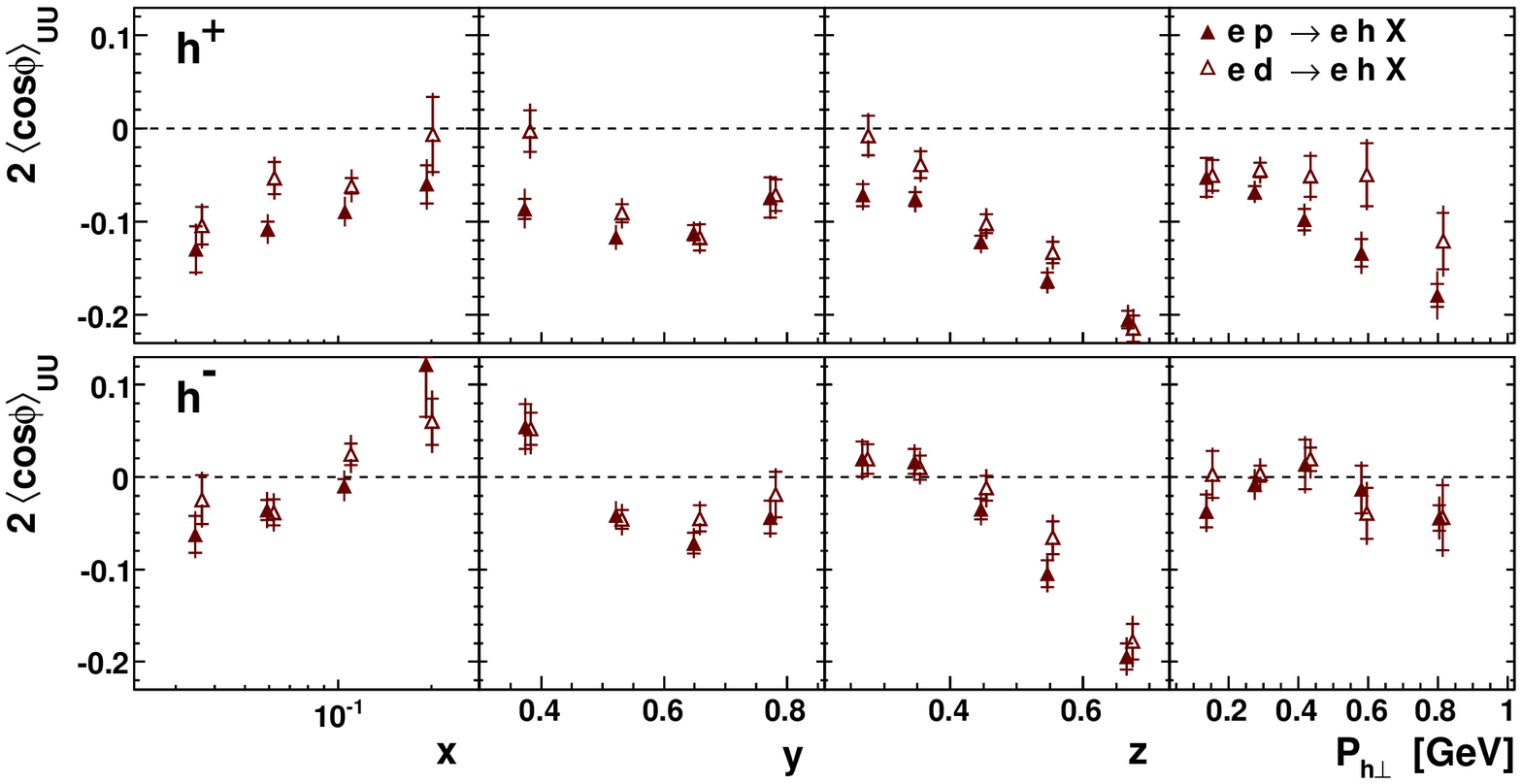}
\caption{As in Fig.~\ref{fig:Pionscos}, but for unidentified charged hadrons.}
\label{fig:Hadronscos}
\end{center}
\end{figure*}

Results for the $\csm$ amplitudes extracted from hydrogen and deuterium data
are shown in Fig.~\ref{fig:Hadronscos}. 
They are found to be negative for both positively and negatively charged hadrons,
but larger in magnitude for the positive hadrons.
Hadrons produced using hydrogen and deuterium targets result in similar
amplitudes, but small differences can be observed for $h^+$ that reflect the behavior of the $\pi^+$ amplitudes.

\subsection{Comparison of amplitudes for various hadron types}
\begin{figure*}[htp]
\begin{center}
\includegraphics[width=0.95\textwidth]{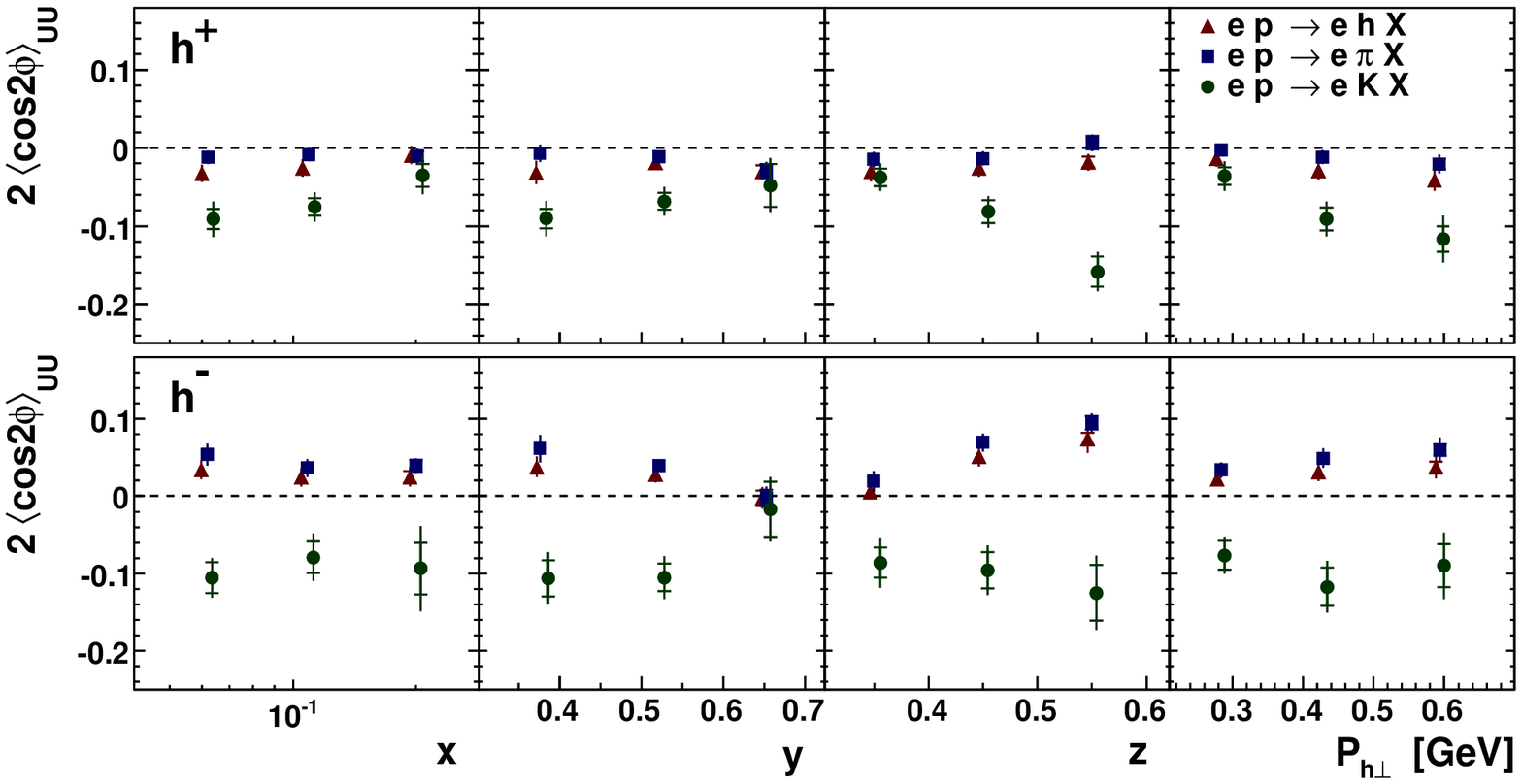}
\caption{$\cstm$ amplitudes from a hydrogen target for positive (upper panels) and negative
  (lower  panels) unidentified hadrons (triangles), pions (squares) and kaons (circles),  integrated over the kinematic range \emph{B} of table~\ref{tab:KineRange}.
 Uncertainties as in Fig.~\ref{fig:Pionscos2}.  Points have been slightly shifted horizontally for visibility.}
\label{fig:Compcos2H}
\end{center}
\end{figure*}

\begin{figure*}[htp]
\begin{center}
\includegraphics[width=0.95\textwidth]{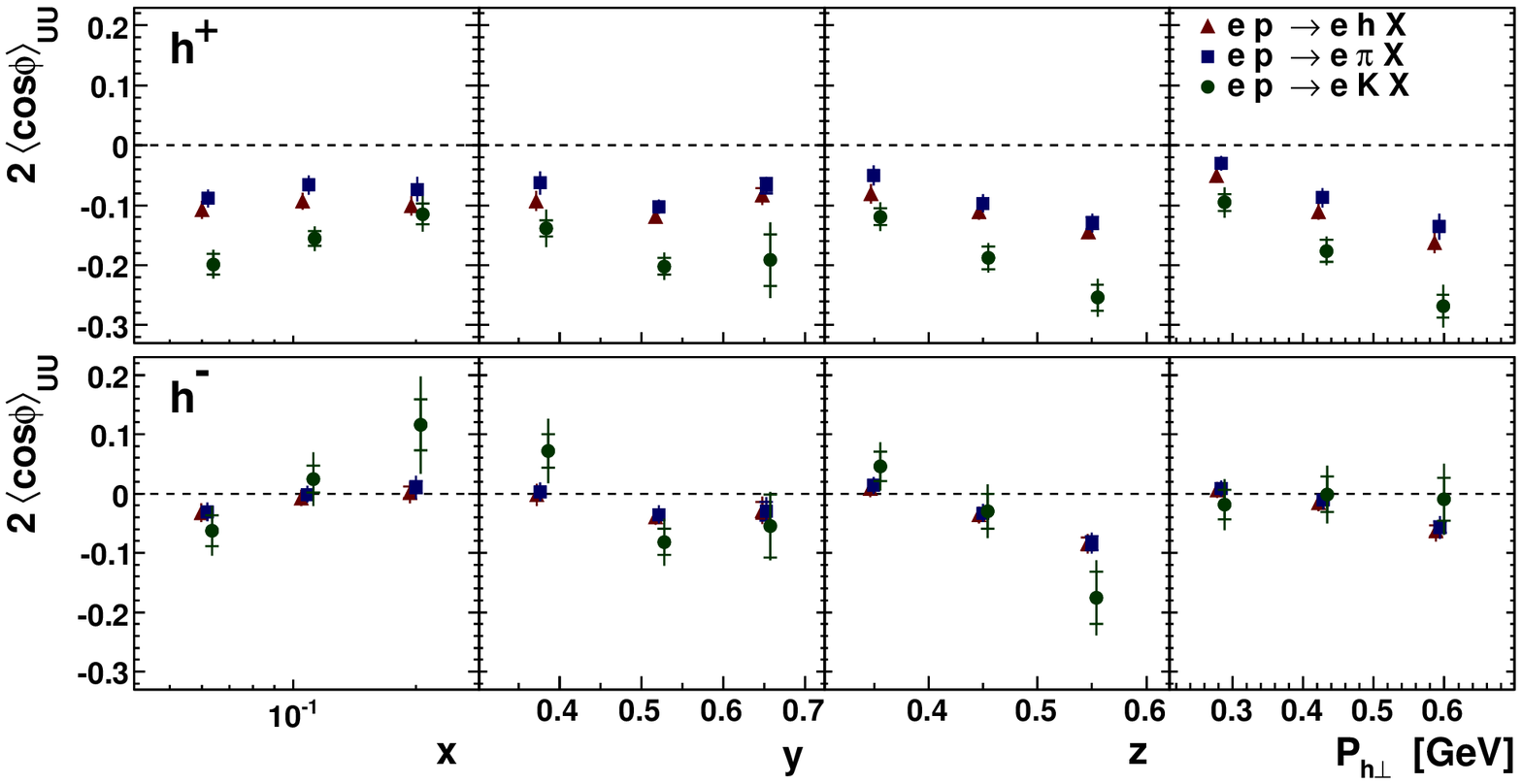}
\caption{$\csm$ amplitudes from a hydrogen target for positive (upper panels) and negative
  (lower  panels) unidentified hadrons (triangles), pions (squares) and kaons (circles),  integrated over the kinematic range \emph{B} of table~\ref{tab:KineRange}.
Uncertainties as in Fig.~\ref{fig:Pionscos2}.  Points have been slightly shifted horizontally for visibility.}
\label{fig:CompcosH}
\end{center}
\end{figure*}

\begin{figure*}[htp]
\begin{center}
\includegraphics[width=0.95\textwidth]{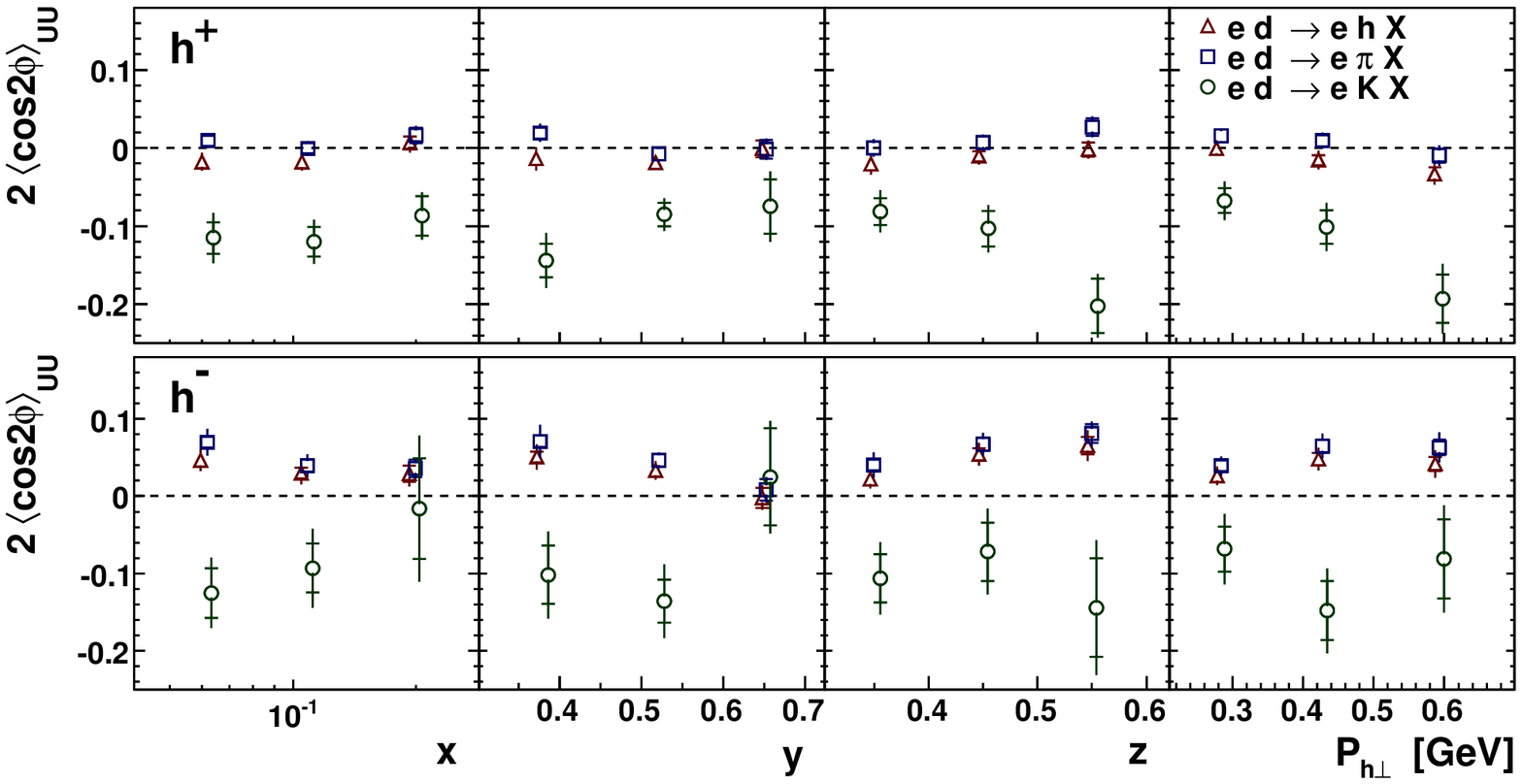}
\caption{As in Fig.~\ref{fig:Compcos2H}, but for a deuterium target.}
\label{fig:Compcos2D}
\end{center}
\end{figure*}

\begin{figure*}[htp]
\begin{center}
\includegraphics[width=0.95\textwidth]{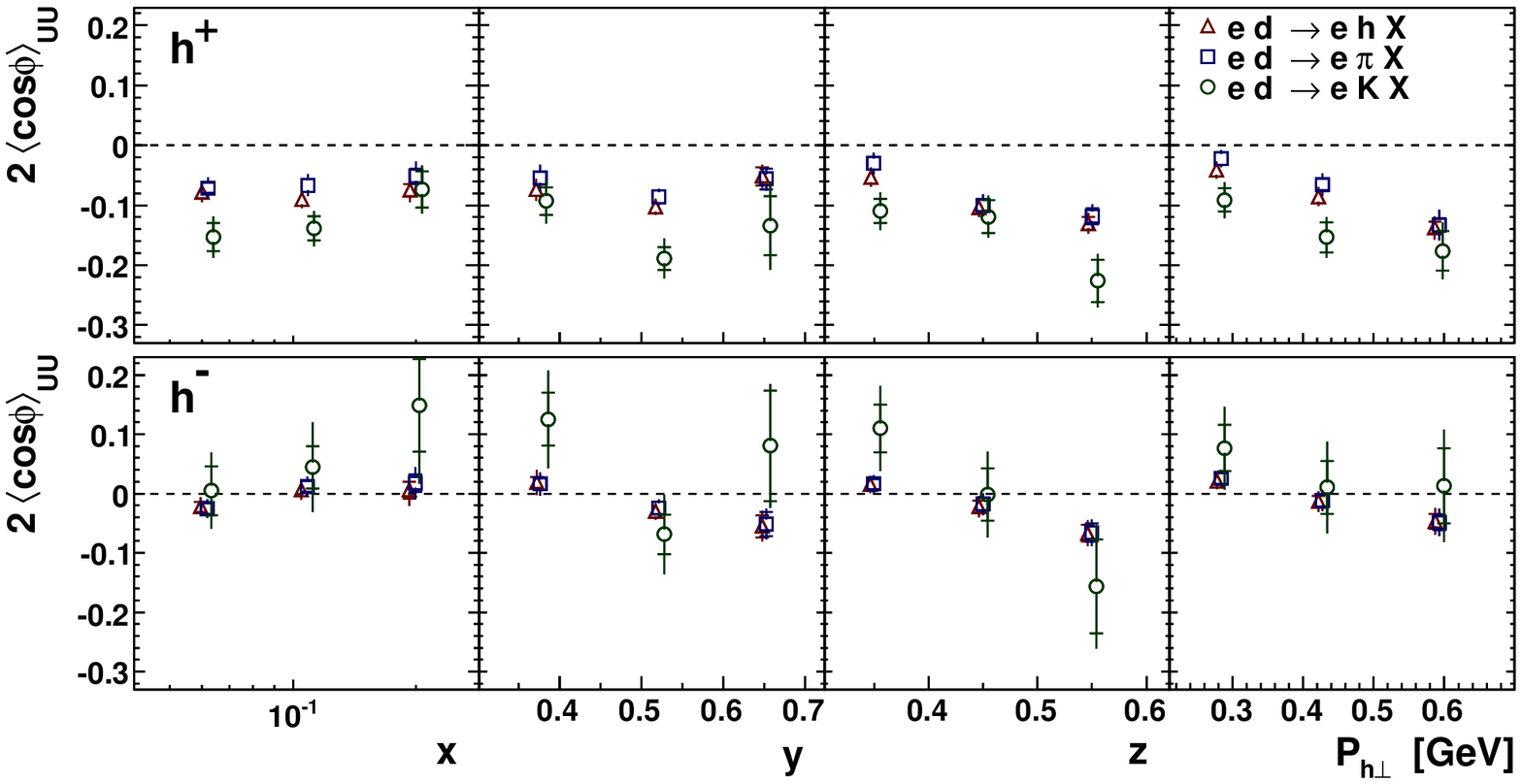}
\caption{As in Fig.~\ref{fig:CompcosH}, but for a deuterium target.}
\label{fig:CompcosD}
\end{center}
\end{figure*}

In order to compare the cosine modulations extracted for the various
hadron types, all samples were projected in the smaller
integration range {\emph B} of table~\ref{tab:KineRange}
and bins were only included if they provided
a measurement in the negative kaon sample produced from a deuterium target,
which is the statistically poorest data sample.
Figures~\ref{fig:Compcos2H}, \ref{fig:CompcosH},
\ref{fig:Compcos2D} and~\ref{fig:CompcosD} show the comparisons 
of the $\cstm$ and $\csm$ amplitudes for
the various hadron types produced on a hydrogen and deuterium target. 
The kaon moments are substantially larger in
magnitude than those of the pions,
with the exception that
the $K^-$ $\cs_\uu$ moments are compatible with those of $\pi^-$,
although their large uncertainties also make them compatible with zero.
The $\cst_\uu$ moments for negative kaons not only have a 
larger magnitude but also the opposite sign as the pion moments.
A magnitude of $K^+$ amplitudes larger than that for $\pi^+$ 
was already observed in the case of the 
amplitudes  measured in transverse-target single-spin asymmetries 
where the Collins \ff\ couples to the transversity \df~\cite{Airapetian:2010ds}.
The large amplitudes for kaons suggest a Collins effect that is larger
for kaons than for pions; in addition, the differences with respect to pions can
arise from a significant role of $strange$ 
quarks in kaon production.
The modulations extracted for unidentified hadrons and pions have similar trends,
although some differences are observed, particularly for the $\cstm$ amplitudes.
The discrepancies between hadrons and pions are generally consistent with the
observed kaon moments.

\subsection{Charge difference}

The systematic uncertainties of results in Figs.~\ref{fig:Pionscos2}--\ref{fig:Hadronscos}
are highly correlated for positive and negative hadrons of the same type, 
as they were measured under the same experimental conditions.
It is therefore useful
to provide the difference between the amplitudes of positive and negative hadrons, 
where many systematic uncertainties cancel. The charge difference provides more 
strict constraints for models, as it accounts for correlated systematics between 
hadrons of the same type, but different charge. In addition some hadron-flavor blind 
contributions to the moment may be suppressed, e.g.,
a \C\  effect as considered so far in most phenomenological approaches. 
In that case, both the $\csm$ and $\cstm$ charge difference amplitudes are expected to have 
an increased sensitivity to the \BMC\  effect.

\begin{figure*}[p]
\begin{center}
\includegraphics[width=0.95\textwidth]{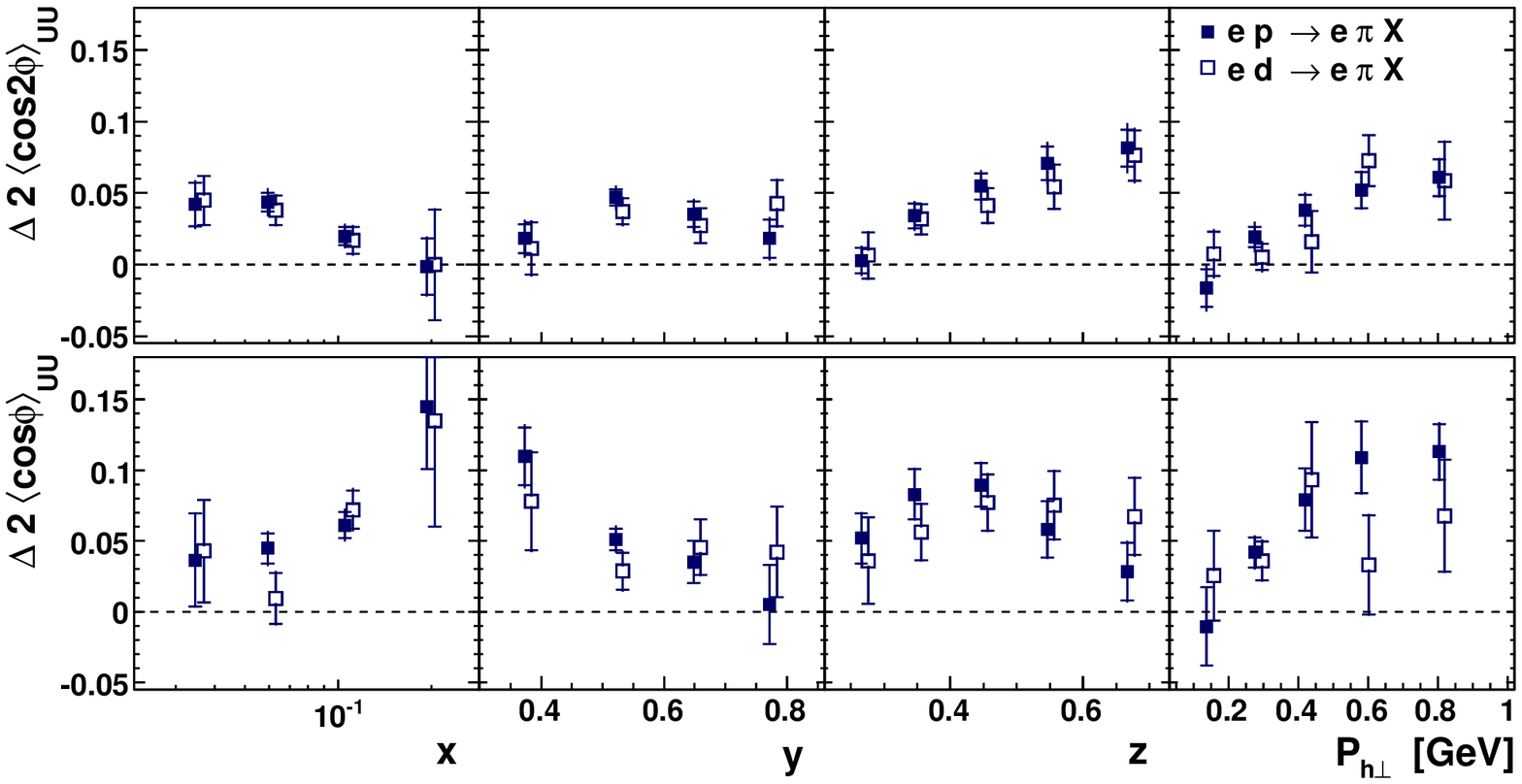}
\caption{Difference between the amplitudes of negative and positive pions within the kinematic range \emph{A} of table~\ref{tab:KineRange}: $2\cst_{\pi^-} - 2\cst_{\pi^+}$ (upper panels), 
$2\cs_{\pi^-} - 2\cs_{\pi^+}$ (lower panels).
Closed and open symbols are for amplitudes extracted from hydrogen and 
deuterium targets, respectively. Points have been slightly shifted horizontally for visibility. 
Uncertainties as in Fig.~\ref{fig:Pionscos2}.}
\label{fig:DiffPi}
\end{center}
\end{figure*}

\begin{figure*}[p]
\begin{center}
\includegraphics[width=0.95\textwidth]{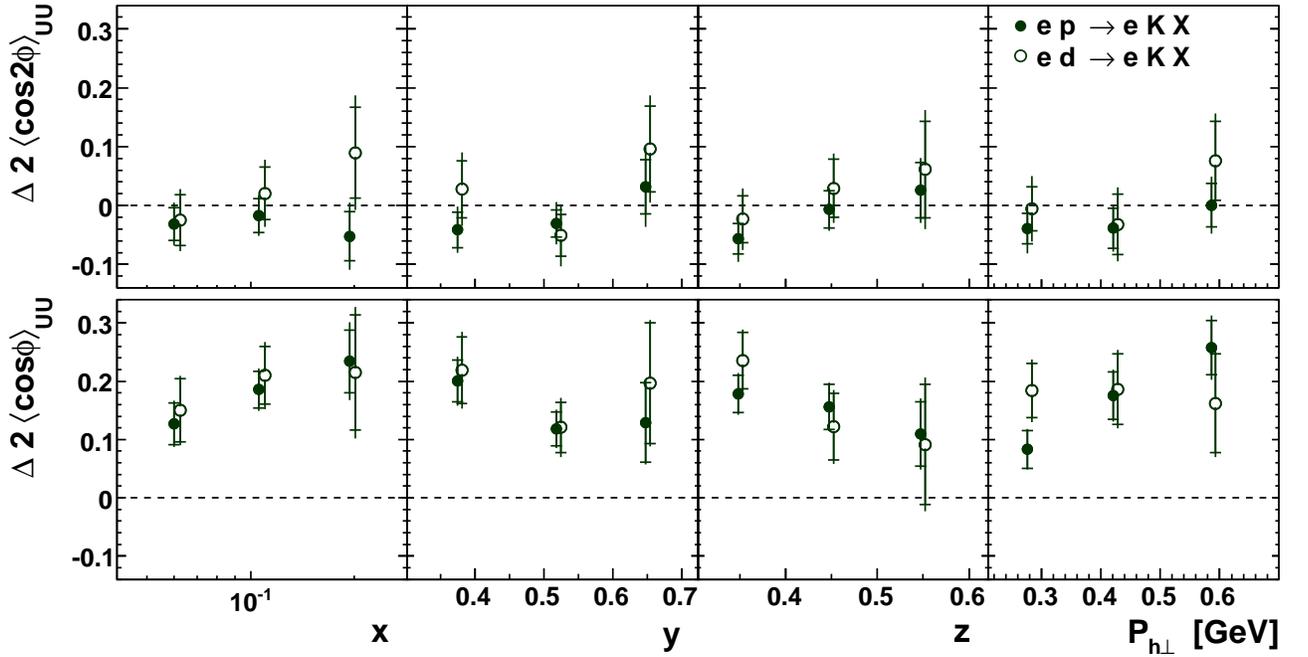}
\caption{As in Fig.~\ref{fig:DiffPi}, but for kaons, and the kinematic range \emph{B} of table~\ref{tab:KineRange}.}
\label{fig:DiffK}
\end{center}
\end{figure*}

\begin{figure*}[t]
\begin{center}
\includegraphics[width=0.95\textwidth]{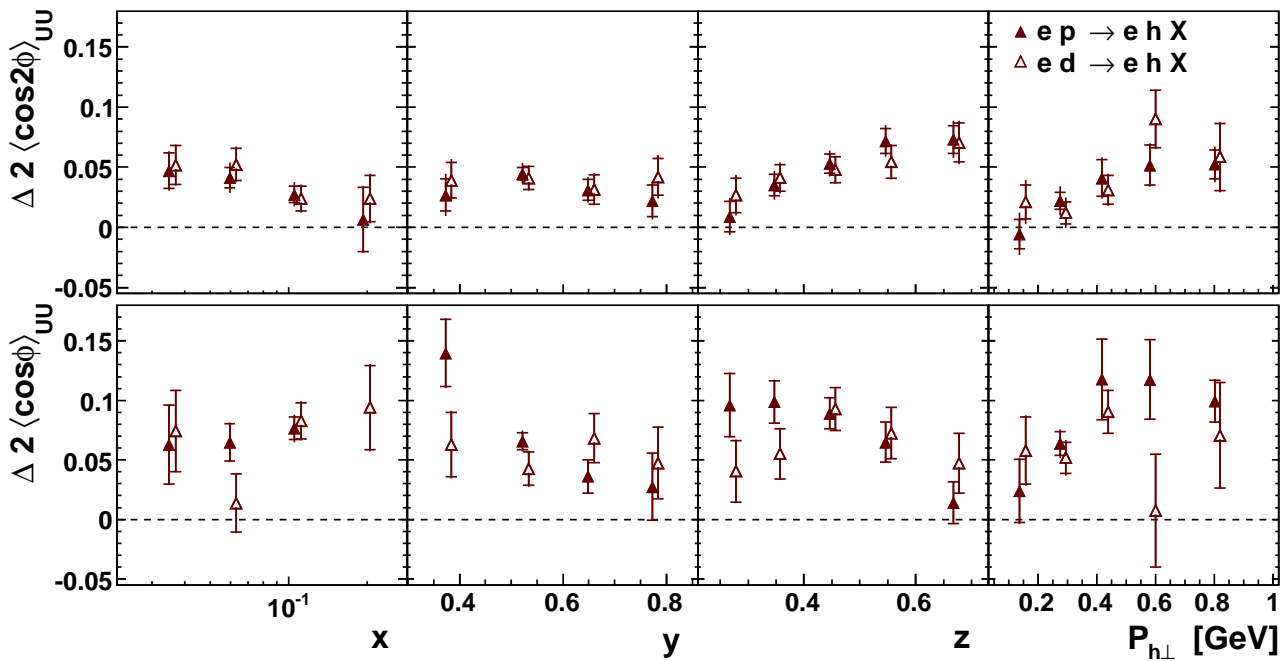}
\caption{As in Fig.~\ref{fig:DiffPi}, but for unidentified hadrons.}
\label{fig:DiffH}
\end{center}
\end{figure*}

For each hadron type the charge difference of the respective amplitudes was evaluated,
and its uncertainty was computed, taking into account the correlations.
The results are shown in Figs.~\ref{fig:DiffPi},~\ref{fig:DiffK}, and~\ref{fig:DiffH},
for pions, kaons, and unidentified hadrons, respectively.
For pions and unidentified hadrons the charge difference is significantly non-zero over nearly 
the entire kinematic range.
For kaons, a large charge difference is observed for the $\csm$ amplitudes,
while the difference for the $\cstm$ amplitudes is compatible with zero.
The different behavior of kaons with respect to pions suggests
an important contribution to the modulations from scattering off 
$strange$ quarks, or, more generally, from scattering off the sea.

\section{Conclusions}

\hermes\ measured fully-differential ($4D$) cosine modulations for charged 
pions, kaons, and unidentified hadrons produced in 
semi-inclusive DIS off unpolarized hydrogen and deuterium targets.
In the TMD framework, these amplitudes can be interpreted 
by a non-zero intrinsic transverse momentum of quarks 
(\C\ effect) and by correlations between the quark's transverse polarization 
and its transverse momentum and 
 the transverse momentum of the produced hadron 
(\BMC\ effect).
However, considering the low average $Q^2$ attainable at \hermes, 
contributions suppressed as $(1/Q)^2$ and higher
may be not negligible.

To date, the cosine modulations presented here 
represent the most complete data set on the subject,
and allow access to flavor-dependent information 
on the internal degrees of freedom of the nucleon.
The extracted $4D$ amplitudes, with their full covariance matrix, provide the maximum information 
from this measurement, and can be used to guide model construction
in a fully differential way.

In addition, the amplitudes have been presented as projections
over specific integration ranges of the four kinematic variables $x$, $y$, $z$,
and $\Pt$.
Significant differences are observed for moments extracted for
oppositely charged pions, interpreted as being due to the convolution of
the \BM\ and the Collins functions.
The extracted amplitudes for pions and kaons show different sizes and kinematic dependences.
In particular, the $\cstm$ amplitudes for kaons are larger than for pions, 
and do not change sign for oppositely charged kaons. 
The former may be due to different features of Collins fragmentation into kaons and pions, 
possibly related to a significant contribution from $strange$ quarks to kaon production.

The amplitudes extracted from hydrogen and deuterium targets 
are found to be similar, but slightly different for positive pions and kaons. 
For pions, a similar size can be due to the
\BM\ distribution functions with the same sign
for $up$ and $down$ quarks. 
For kaons this can be due to a similar contribution from $up$ and $down$ quarks
along with a similar $strange$ sea distribution in protons and neutrons.
The slight differences for hydrogen and deuterium targets for positive pions and kaons
might be related to a slightly different magnitude of the \BM\ functions for the different quark types.
The difference of moments between positively and negatively charged hadrons is not 
compatible with zero for all hadron types except for the $\cstm$ kaons. 
The different behavior of kaons with respect to pions suggests
a significant contribution to the modulations 
from scattering off 
$strange$ quarks, or, more generally, from scattering off the sea, 
or from fragmentation of light quarks into kaons.

\acknowledgements
We gratefully acknowledge the \desy\ management for its support and the staff
at \desy\ and the collaborating institutions for their significant effort.
This work was supported by 
the Ministry of Economy and the Ministry of Education and Science of Armenia;
the FWO-Flanders and IWT, Belgium;
the Natural Sciences and Engineering Research Council of Canada;
the National Natural Science Foundation of China;
the Alexander von Humboldt Stiftung,
the German Bundesministerium f\"ur Bildung und Forschung (BMBF), and
the Deutsche Forschungsgemeinschaft (DFG);
the Italian Istituto Nazionale di Fisica Nucleare (INFN);
the MEXT, JSPS, and G-COE of Japan;
the Dutch Foundation for Fundamenteel Onderzoek der Materie (FOM);
the Russian Academy of Science and the Russian Federal Agency for 
Science and Innovations;
the U.K.~Engineering and Physical Sciences Research Council, 
the Science and Technology Facilities Council,
and the Scottish Universities Physics Alliance;
the U.S.~Department of Energy (DOE) and the National Science Foundation (NSF);
the Basque Foundation for Science (IKERBASQUE) and the UPV/EHU under program UFI 11/55;
and the European Community Research Infrastructure Integrating Activity
under the FP7 "Study of strongly interacting matter (HadronPhysics2, Grant
Agreement number 227431)".

%% file: appendix.tex
\section{Analysis methods}\label{sec:Appdx}

In this section some technical details particular to this analysis are discussed.
Section~\ref{sec:FoldFit} includes the details of the fully
differential unfolding and fitting procedure.
Section~\ref{sec:Param} describes the extraction
of a $4D$ model of the measured cosine modulations from the data.

\subsection{Five-dimensional unfolding and fitting}\label{sec:FoldFit}
As described in section~\ref{sec:procedure}, the measured yields
are simultaneously unfolded (i.e., corrected for acceptance, smearing and QED
radiative effects) and fit by minimizing the $\chi^2$
in Eq.~\ref{eq:FoldFit}.
In Equation~\ref{eq:FoldFit}, $\sigma^{\text{data}}$ is a vector and $S$ and $C$ are square matrices, 
all of dimension of the number of bins ($5*5*6*6*12=10800$, see Table~\ref{tab:Bins}).
The results vector $\beta$ contains the three fit parameters (${\mathcal A}$, ${\mathcal B}$, and ${\mathcal C}$, see Eq.~\ref{eqn:ABC}) 
for each ($x$, $y$, $z$, $P_{h\perp}$) bin and therefore is of dimension
$(5*5*6*6)*3 = 2700$.  
The product $X\beta$ gives the fit function of Eq.~\ref{eqn:ABC} in each of the 10800 bins and so $X$ is 10800 by 2700.  
Each row contains elements equal to 1, $\cos\langle\ph\rangle$, and $\cos2 \langle\ph\rangle$ for that bin; the remaining elements are 0.
The result is that $X$ is block diagonal, with blocks of dimension 12x3.

The covariance $C$ includes the 
sum of three sources of statistical uncertainties:
the precision of the measured yields, the precision of the \MC\ 
used for the background subtraction and the precision of the \MC\ used 
to construct the smearing matrix\footnote{It can be shown mathematically that 
if the results are first unfolded (including only the first two contributions), 
and then fit (including the additional uncertainty due to S), 
an identical term will appear in the calculation of the covariance of the fit parameters.}.
These three sources are accounted for in the covariance following the
standard uncertainty propagation.
The first two terms are the diagonal covariances of $\sigma_\uu^{\text{raw}}$ (the raw yields), 
and $\sigma_\uu^{\text{backgr}}$ (the background yields).
Together, these give the uncertainty of background-subtracted 
yields, i.e., of $\sigma_\uu^{\text{data}} = \sigma_\uu^{\text{raw}}-\sigma_\uu^{\text{backgr}}$.
The third term accounts for the
propagation of the smearing matrix covariance $C_S$ through
the full unfolding and fitting procedure.
$C_S$ is calculated from the 
statistical uncertainty of the migration matrix and the Born-level 
simulated yields used to calculate the smearing matrix.
This uncertainty contributes an additive term in $C$ of the form
\begin{equation}
C^{\text{smear}} = \sigma_\uu^{\text{unf}}\; C_S\; {\sigma_\uu^{\text{unf}}}^{\,T},
\end{equation}
where $\sigma_\uu^{\text{unf}}$ is the unfolded (Born-level) yield vector,
which is calculated by correcting the measured yields for smearing and background.
This third contribution to the covariance is small, 
as the \MC\ productions contain 
approximately $20$ times as many events as the data productions.

The $\chi^2$ defined in Eq.~\ref{eq:FoldFit} was minimized with 
respect to the vector of parameters $\beta$ by means 
of linear regression,
producing the parameters
\begin{align}
\beta = &(X^TS^T{C}^{-1}SX)^{-1}X^TS^T {C}^{-1}\sigma_\uu^{\text{data}},
\end{align}
along with their covariance
\begin{align}
C_{\beta} = &(X^TS^T{C}^{-1}SX)^{-1}.
\end{align}
Here, a generalized procedure was used that unfolds separately each data set 
with its own smearing matrix, and then simultaneously fits
the unfolded yields from the various years\footnote{Mathematically this is 
equivalent to taking the weighted average of the 
fit parameters from various data sets before calculating the moments 
(which correspond to the ratio of the fit parameters).}.
For this purpose, a \emph{super matrix} form of Eq.~\ref{eq:FoldFit} was defined,
\be\label{eq:superchi}
\chi^2= ({\bf\boldsymbol\sigma_\uu^{S}} - {\bf S^{S}} X\beta)^T {\bf
  C^{-1}_{S}} ({\bf\boldsymbol \sigma_\uu^{S}} - {\bf S^{S}}X\beta),
\ee
where ${\bf\boldsymbol\sigma_\uu^{S}}$ is a super vector that includes
the $\sigma_\uu^{\text{data},d}$ for all data sets $d=1,..,n$
\begin{align}
{\bf\boldsymbol\sigma_\uu^{S}}= \left(\! \begin{array}{c}
    \sigma_\uu^{\text{data},1} \\ .. \\ \sigma_\uu^{\text{data},n} \end{array} \! \right).
\end{align}

The super matrices ${\bf S^{S}}$ and ${\bf C_{S}}$ include
respectively the smearing matrices and their covariances for various data sets:
\begin{align}
{\bf S^{S}}  = \left(\! \begin{array}{c} S_1 \\ .. \\ S_n \end{array} \! \right), \quad
{\bf C_{S}}  = \left(\! \begin{array}{ccc}  C^{-1}_{1} & 0 & 0 \\ 0 &
    .. & 0 \\ 0 &0 &  C^{-1}_{n} \end{array} \! \right).
\end{align}
Since ${\bf C_{S}}$ is a block-diagonal matrix,
equation~\ref{eq:superchi} gives
\begin{align}
 \beta = &\left\{ X^T  \left( \sum_{d=1}^n S^T_d C^{-1}_{d} S_d  \right) X \right\}^{-1} \notag\\ 
 & X^T   \left( \sum_{d=1}^n S^T_d C^{-1}_{d} \sigma_\uu^{\text{data},d} \right) 
\end{align}
and covariance 
\begin{align}
&C_\beta = \left\{ X^T  \left( \sum_{d=1}^n S^T_d C^{-1}_{d} S_d  \right) X \right\}^{-1}.
\end{align}

\subsection{$4D$ model of the extracted moments}\label{sec:Param}
Several of the systematic tests described in section~\ref{sec:sys} require a \MC\ production that
includes azimuthal modulations consistent with those found in the data.  
To facilitate this, the fully differential final results are fit to a four-dimensional function,
which is then used to alter the underlying distribution in an azimuthally 
independent \pythia\ \MC\ production. 
The fit function used has $38$ parameters, $19$ for each modulation ($\cstm$, $\csm$),
and is of the form:
\begin{align}
        f= &A_1 + A_2x + A_3y + A_4z + A_5\Pt\nonumber \\
           &+ A_6x^2  + A_7z^2 + A_8\Pt^2 + A_9xz\nonumber \\
           &+ A_{10}x\Pt + A_{11}z\Pt + A_{12}y \Pt\nonumber  \\
           &+ A_{13} yx + A_{14}yz  + A_{15}y^2 \nonumber\\
           &+ A_{16}x^3  + A_{17}z^3 + A_{18}\Pt^3 + A_{19}y^3. 
\end{align}
Several other functional forms were also tested and gave compatible
results. This procedure was used to extract one model separately
for hydrogen and deuterium targets, and for each particle type and charge: 
pions, kaons, and unidentified hadrons.

\section{Hadron Identification}
\label{sec:EVT}
\label{sec:RICHalgorithm}

The \hermes\ dual-radiator ring-imaging Cherenkov (RICH)
detector is described in detail in Ref.~\cite{Akopov:2000qi}.
In that article the indirect ray tracing (IRT) particle identification algorithm
is presented.
In addition, an alternative method, the direct ray tracing (DRT) algorithm
was developed, which is described in Ref.~\cite{Cisbani1999366}.
The IRT algorithm calculates an expected Cherenkov angle and compares this to the observed photons.
The DRT algorithm performs a simulation, generating an expected photon pattern for a given particle hypothesis, which is then compared to the observed pattern.
Because DRT performs a full simulation it is more accurate than IRT, at the cost of increased computing time.

Here, a new method, EVT, is presented.
This event-level algorithm can more effectively identify tracks in semi-inclusive events 
where rings from several tracks may overlap.

\subsection{The EVT algorithm}
The DRT algorithm generates a simulated photomultiplier tube (PMT) hit pattern for each radiator, based on known track kinematics and a particle-type hypothesis.
Hit patterns are generated for each particle type hypothesis (pion, kaon, proton) and 
the likelihood of each hypothesis is calculated by comparing the simulated hit pattern to the hit distribution observed.
Due to computing constraints, DRT was initially only used on a subset of particle tracks, 
and the event-level sum over tracks shown in the equations of section 3.1 
of Ref.~\cite{Cisbani1999366} was not computed.
Advances in CPU power in recent years made it possible for the DRT method to be run on all tracks in all events, but
the event-level consideration of all tracks in an event was not implemented in the software.

The EVT method implements the full event-level identification algorithm described in Ref.~\cite{Cisbani1999366}.
Simulated hit patterns for each track are 
combined for all permutations of particle hypotheses to form a set of predicted hit patterns for the event.
The likelihood ($L^{H}$) of an event hypothesis $H$ is given by the observed hit pattern ($C_{\mathrm{PMT}}(i)$)
and the probability of hit ($P^{(H)}_{\mathrm{PMT}}(i)$),
%
%
%
\begin{multline} \label{eqn:LLevt}
L^{H} = \sum_{i} \log\left[ P^{H}_{\mathrm{PMT}}(i) C_{\mathrm{PMT}}(i) \right. \\
\left. +\,\bar{P}^{H}_{\mathrm{PMT}}(i)(1 - C_{\mathrm{PMT}}(i))\right],
\end{multline}
where the probability of no hit is simply $\bar{P}^{H}_{\mathrm{PMT}}(i) = 1 - P^{H}_{\mathrm{PMT}}(i)$.
The observed hit pattern $C_{\mathrm{PMT}}(i)$ is 1 if PMT $i$ is hit, and 0 otherwise.
An event hypothesis $H$ is a set of particle type hypotheses, one for each track in the event.  
Given $T$ tracks, each with $h$ (=3, pions, kaons, proton) possible hypotheses, there are a total of $h^T$ event hypotheses $H$.  
The hypothesis for track $t$ given the event hypothesis $H$ is $H_t$.
The simulated hit pattern of track $t$, of particle type $H_t$, produced from radiator $r$ is given by $N^{(H_t,t,r)}(i)$.
The probability $P^{H}_{\mathrm{PMT}}(i)$ of a hit given the event hypothesis $H$, is 
%
%
\begin{align}
& P^H_{\mathrm{PMT}}(i)  =  \nonumber \\
& \quad 1- \exp\!\left(
- \sum_{r,t} \left[ \frac{ N^{(H_t,t,r)}(i)} {\sum_i N^{(H_t,t,r)}(i)}  n^{(H_t,t,r)} \right] - B(i) \right), \label{eqn:Probevt}
\end{align}
where the sum is over the hits from both radiators ($r$) and all the tracks ($t$) in the event.
An unphysically high number of hits is simulated for each radiator ($\sum_i N^{(H_t,t,r)}(i)=360$) 
to construct a smooth distribution of the expected hits.
The simulated hit pattern($N^{(H_t,t,r)}(i)$) is then normalized ( $\frac{ n^{(H_t,t,r)}} {\sum_i N^{(H_t,t,r)}(i)}$ )
to the number of expected PMT hits for the given particle type, track kinematics, and radiator.  The total number of expected hits, $n^{(H_t,t,r)}$, is typically 0-10 hits.
The $B(i)$ term is included to take into account physical and experimental backgrounds; 
see section~\ref{sec:backgrounds} for more details.

After the likelihoods $L^{H}$ are computed, the most likely is chosen and the particle type of each track in the event is given by $H_t$.

\subsection{Backgrounds} \label{sec:backgrounds}
The background term $B(i)$ was investigated by counting the average number of hits in each PMT in the absence of 
tracks in that detector half.  Average background values of ~0.005 were observed in the data,
with a structure that shows the highest values nearest to the beam, indicating that the origin was not only 
electronic noise. Event displays showed that these ``background'' events that have no associated tracks in fact 
show a ring structure in the RICH.
Extracting the same quantity from Monte Carlo simulation showed similar results, confirming that these ``background'' rings 
are from physical particles not tracked by the spectrometer.
They are caused by high-energy photons from $\pi^{0}$ decays and bremsstrahlung of the scattered beam lepton, 
producing $e^+e^-$ pairs in front of the RICH.

While the $B(i)$ term can only provide an average treatment of such untracked rings, it is the simplest way to 
take this physical background into account using the existing algorithm.  \MC\ tests
 using $B(i)$ determined from data showed that it led to more efficient particle identification 
with less contamination compared to the constant value of 0.0001 (as was used in the past).

Extracting $B(i)$ from various subsets of the data showed that $B(i)$ is relatively insensitive to the type of target gas. 
However, it is affected by the gas density, showing higher values for targets with a higher density.  
When applied to simulated data, a larger $B(i)$ favors identification (both correctly and incorrectly) as a (anti)proton 
while a smaller $B(i)$ favors identification (both correctly and incorrectly) as a kaon.  
In each case there is a trade off between efficiency and contamination.  
Since the overall flux of pions is largest, it is relatively unaffected by the change while the very small antiproton flux 
shows relatively large changes when using a different $B(i)$.  
However, the use of different background files had a negligible affect on the cosine modulations extracted here.
$B(i)$ was extracted from the unpolarized data for each data year and used for the RICH algorithm applied to all of the data from that year.

%% file: draft78.bbl
\begin{thebibliography}{85}
\expandafter\ifx\csname natexlab\endcsname\relax\def\natexlab#1{#1}\fi
\expandafter\ifx\csname bibnamefont\endcsname\relax
  \def\bibnamefont#1{#1}\fi
\expandafter\ifx\csname bibfnamefont\endcsname\relax
  \def\bibfnamefont#1{#1}\fi
\expandafter\ifx\csname citenamefont\endcsname\relax
  \def\citenamefont#1{#1}\fi
\expandafter\ifx\csname url\endcsname\relax
  \def\url#1{\texttt{#1}}\fi
\expandafter\ifx\csname urlprefix\endcsname\relax\def\urlprefix{URL }\fi
\providecommand{\bibinfo}[2]{#2}
\providecommand{\eprint}[2][]{\url{#2}}

\bibitem[{\citenamefont{Bjorken and Paschos}(1969)}]{Bjorken:1969ja}
\bibinfo{author}{\bibfnamefont{J.~D.} \bibnamefont{Bjorken}} \bibnamefont{and}
  \bibinfo{author}{\bibfnamefont{E.~A.} \bibnamefont{Paschos}},
  \bibinfo{journal}{Phys. Rev.} \textbf{\bibinfo{volume}{185}},
  \bibinfo{pages}{1975} (\bibinfo{year}{1969}).

\bibitem[{\citenamefont{Feynman}(1969)}]{Feynman:1969ej}
\bibinfo{author}{\bibfnamefont{R.~P.} \bibnamefont{Feynman}},
  \bibinfo{journal}{Phys. Rev. Lett.} \textbf{\bibinfo{volume}{23}},
  \bibinfo{pages}{1415} (\bibinfo{year}{1969}).

\bibitem[{\citenamefont{de~Florian et~al.}(2009)\citenamefont{de~Florian,
  Sassot, Stratmann, and Vogelsang}}]{deFlorian:2009vb}
\bibinfo{author}{\bibfnamefont{D.}~\bibnamefont{de~Florian}},
  \bibinfo{author}{\bibfnamefont{R.}~\bibnamefont{Sassot}},
  \bibinfo{author}{\bibfnamefont{M.}~\bibnamefont{Stratmann}},
  \bibnamefont{and}
  \bibinfo{author}{\bibfnamefont{W.}~\bibnamefont{Vogelsang}},
  \bibinfo{journal}{Phys. Rev.} \textbf{\bibinfo{volume}{D80}},
  \bibinfo{pages}{034030} (\bibinfo{year}{2009}).

\bibitem[{\citenamefont{Aaron et~al.}(2010)}]{Aaron:2009wt}
\bibinfo{author}{\bibfnamefont{F.~D.} \bibnamefont{Aaron}} \bibnamefont{et~al.}
  (\bibinfo{collaboration}{H1 and ZEUS}), \bibinfo{journal}{JHEP}
  \textbf{\bibinfo{volume}{1001}}, \bibinfo{pages}{109} (\bibinfo{year}{2010}).

\bibitem[{\citenamefont{Pumplin et~al.}(2002)}]{Pumplin:2002vw}
\bibinfo{author}{\bibfnamefont{J.}~\bibnamefont{Pumplin}} \bibnamefont{et~al.},
  \bibinfo{journal}{JHEP} \textbf{\bibinfo{volume}{0207}}, \bibinfo{pages}{012}
  (\bibinfo{year}{2002}).

\bibitem[{\citenamefont{Mulders and Tangerman}(1996)}]{Mulders:1995dh}
\bibinfo{author}{\bibfnamefont{P.~J.} \bibnamefont{Mulders}} \bibnamefont{and}
  \bibinfo{author}{\bibfnamefont{R.~D.} \bibnamefont{Tangerman}},
  \bibinfo{journal}{Nucl. Phys.} \textbf{\bibinfo{volume}{B461}},
  \bibinfo{pages}{197} (\bibinfo{year}{1996}).

\bibitem[{\citenamefont{Collins and Soper}(1981)}]{Collins:1981uk}
\bibinfo{author}{\bibfnamefont{J.~C.} \bibnamefont{Collins}} \bibnamefont{and}
  \bibinfo{author}{\bibfnamefont{D.~E.} \bibnamefont{Soper}},
  \bibinfo{journal}{Nucl. Phys.} \textbf{\bibinfo{volume}{B193}},
  \bibinfo{pages}{381} (\bibinfo{year}{1981}).

\bibitem[{\citenamefont{Boer and Mulders}(1998)}]{Boer:1997nt}
\bibinfo{author}{\bibfnamefont{D.}~\bibnamefont{Boer}} \bibnamefont{and}
  \bibinfo{author}{\bibfnamefont{P.~J.} \bibnamefont{Mulders}},
  \bibinfo{journal}{Phys. Rev.} \textbf{\bibinfo{volume}{D57}},
  \bibinfo{pages}{5780} (\bibinfo{year}{1998}).

\bibitem[{\citenamefont{Ji et~al.}(2004)\citenamefont{Ji, Ma, and
  Yuan}}]{Ji:2004xq}
\bibinfo{author}{\bibfnamefont{X.}~\bibnamefont{Ji}},
  \bibinfo{author}{\bibfnamefont{J.-P.} \bibnamefont{Ma}}, \bibnamefont{and}
  \bibinfo{author}{\bibfnamefont{F.}~\bibnamefont{Yuan}},
  \bibinfo{journal}{Phys. Lett.} \textbf{\bibinfo{volume}{B597}},
  \bibinfo{pages}{299} (\bibinfo{year}{2004}).

\bibitem[{\citenamefont{Ji et~al.}(2005)\citenamefont{Ji, Ma, and
  Yuan}}]{Ji:2004wu}
\bibinfo{author}{\bibfnamefont{X.}~\bibnamefont{Ji}},
  \bibinfo{author}{\bibfnamefont{J.-P.} \bibnamefont{Ma}}, \bibnamefont{and}
  \bibinfo{author}{\bibfnamefont{F.}~\bibnamefont{Yuan}},
  \bibinfo{journal}{Phys. Rev.} \textbf{\bibinfo{volume}{D71}},
  \bibinfo{pages}{034005} (\bibinfo{year}{2005}).

\bibitem[{\citenamefont{Collins}(2011)}]{Collins:book}
\bibinfo{author}{\bibfnamefont{J.~C.} \bibnamefont{Collins}},
  \emph{\bibinfo{title}{Foundations of Perturbative QCD}}
  (\bibinfo{publisher}{Cambridge University Press}, \bibinfo{address}{London},
  \bibinfo{year}{2011}).

\bibitem[{\citenamefont{Feynman}(1998)}]{Feynman72}
\bibinfo{author}{\bibfnamefont{R.}~\bibnamefont{Feynman}},
  \emph{\bibinfo{title}{Photon-hadron interactions}}, Advanced book classics
  (\bibinfo{publisher}{Addison-Wesley}, \bibinfo{year}{1998}).

\bibitem[{\citenamefont{Ravndal}(1973)}]{Ravndal73}
\bibinfo{author}{\bibfnamefont{F.}~\bibnamefont{Ravndal}},
  \bibinfo{journal}{Physics Letters} \textbf{\bibinfo{volume}{B43}},
  \bibinfo{pages}{301 } (\bibinfo{year}{1973}).

\bibitem[{\citenamefont{Kingsley}(1974)}]{Kingsley74}
\bibinfo{author}{\bibfnamefont{R.~L.} \bibnamefont{Kingsley}},
  \bibinfo{journal}{Phys. Rev.} \textbf{\bibinfo{volume}{D10}},
  \bibinfo{pages}{1580} (\bibinfo{year}{1974}).

\bibitem[{\citenamefont{Kotzinian}(1995)}]{Kotzinian:1994dv}
\bibinfo{author}{\bibfnamefont{A.}~\bibnamefont{Kotzinian}},
  \emph{\bibinfo{title}{{New quark distributions and semiinclusive
  electroproduction on the polarized nucleons}}} (\bibinfo{year}{1995}).

\bibitem[{\citenamefont{Tangerman and Mulders}(1995)}]{Tangerman:1995hw}
\bibinfo{author}{\bibfnamefont{R.}~\bibnamefont{Tangerman}} \bibnamefont{and}
  \bibinfo{author}{\bibfnamefont{P.}~\bibnamefont{Mulders}},
  \bibinfo{journal}{Phys.Lett.} \textbf{\bibinfo{volume}{B352}},
  \bibinfo{pages}{129} (\bibinfo{year}{1995}).

\bibitem[{\citenamefont{Cahn}(1978)}]{Cahn:1978se}
\bibinfo{author}{\bibfnamefont{R.~N.} \bibnamefont{Cahn}},
  \bibinfo{journal}{Phys. Lett.} \textbf{\bibinfo{volume}{B78}},
  \bibinfo{pages}{269} (\bibinfo{year}{1978}).

\bibitem[{\citenamefont{Cahn}(1989)}]{Cahn:1989yf}
\bibinfo{author}{\bibfnamefont{R.~N.} \bibnamefont{Cahn}},
  \bibinfo{journal}{Phys. Rev.} \textbf{\bibinfo{volume}{D40}},
  \bibinfo{pages}{3107} (\bibinfo{year}{1989}).

\bibitem[{\citenamefont{Sivers}(1990)}]{Sivers:1990cc}
\bibinfo{author}{\bibfnamefont{D.~W.} \bibnamefont{Sivers}},
  \bibinfo{journal}{Phys. Rev.} \textbf{\bibinfo{volume}{D41}},
  \bibinfo{pages}{83} (\bibinfo{year}{1990}).

\bibitem[{\citenamefont{Sivers}(1991)}]{Sivers:1991fh}
\bibinfo{author}{\bibfnamefont{D.~W.} \bibnamefont{Sivers}},
  \bibinfo{journal}{Phys. Rev.} \textbf{\bibinfo{volume}{D43}},
  \bibinfo{pages}{261} (\bibinfo{year}{1991}).

\bibitem[{\citenamefont{Collins}(1993)}]{Collins:1993kk}
\bibinfo{author}{\bibfnamefont{J.~C.} \bibnamefont{Collins}},
  \bibinfo{journal}{Nucl. Phys.} \textbf{\bibinfo{volume}{B396}},
  \bibinfo{pages}{161} (\bibinfo{year}{1993}).

\bibitem[{\citenamefont{Brodsky et~al.}(2002)\citenamefont{Brodsky, Hwang, and
  Schmidt}}]{Brodsky:2002cx}
\bibinfo{author}{\bibfnamefont{S.~J.} \bibnamefont{Brodsky}},
  \bibinfo{author}{\bibfnamefont{D.~S.} \bibnamefont{Hwang}}, \bibnamefont{and}
  \bibinfo{author}{\bibfnamefont{I.}~\bibnamefont{Schmidt}},
  \bibinfo{journal}{Phys. Lett.} \textbf{\bibinfo{volume}{B530}},
  \bibinfo{pages}{99} (\bibinfo{year}{2002}).

\bibitem[{\citenamefont{Collins}(2002)}]{Collins:2002kn}
\bibinfo{author}{\bibfnamefont{J.~C.} \bibnamefont{Collins}},
  \bibinfo{journal}{Phys. Lett.} \textbf{\bibinfo{volume}{B536}},
  \bibinfo{pages}{43} (\bibinfo{year}{2002}).

\bibitem[{\citenamefont{Belitsky et~al.}(2003)\citenamefont{Belitsky, Ji, and
  Yuan}}]{Belitsky:2002sm}
\bibinfo{author}{\bibfnamefont{A.~V.} \bibnamefont{Belitsky}},
  \bibinfo{author}{\bibfnamefont{X.}~\bibnamefont{Ji}}, \bibnamefont{and}
  \bibinfo{author}{\bibfnamefont{F.}~\bibnamefont{Yuan}},
  \bibinfo{journal}{Nucl. Phys.} \textbf{\bibinfo{volume}{B656}},
  \bibinfo{pages}{165} (\bibinfo{year}{2003}).

\bibitem[{\citenamefont{Airapetian
  et~al.}(2010{\natexlab{a}})}]{Airapetian:2009wj}
\bibinfo{author}{\bibfnamefont{A.}~\bibnamefont{Airapetian}}
  \bibnamefont{et~al.} (\bibinfo{collaboration}{HERMES Collaboration}),
  \bibinfo{journal}{Phys.~Lett.} \textbf{\bibinfo{volume}{B682}},
  \bibinfo{pages}{351} (\bibinfo{year}{2010}{\natexlab{a}}).

\bibitem[{\citenamefont{Cheng and Zee}(1972)}]{ChengZee72}
\bibinfo{author}{\bibfnamefont{T.~P.} \bibnamefont{Cheng}} \bibnamefont{and}
  \bibinfo{author}{\bibfnamefont{A.}~\bibnamefont{Zee}},
  \bibinfo{journal}{Phys. Rev. D} \textbf{\bibinfo{volume}{6}},
  \bibinfo{pages}{885} (\bibinfo{year}{1972}).

\bibitem[{\citenamefont{Bacchetta et~al.}(2007)}]{Bacchetta:2006tn}
\bibinfo{author}{\bibfnamefont{A.}~\bibnamefont{Bacchetta}}
  \bibnamefont{et~al.}, \bibinfo{journal}{JHEP}
  \textbf{\bibinfo{volume}{0702}}, \bibinfo{pages}{093} (\bibinfo{year}{2007}).

\bibitem[{\citenamefont{Bacchetta et~al.}(2004)\citenamefont{Bacchetta,
  D'Alesio, Diehl, and Miller}}]{Bacchetta:2004jz}
\bibinfo{author}{\bibfnamefont{A.}~\bibnamefont{Bacchetta}},
  \bibinfo{author}{\bibfnamefont{U.}~\bibnamefont{D'Alesio}},
  \bibinfo{author}{\bibfnamefont{M.}~\bibnamefont{Diehl}}, \bibnamefont{and}
  \bibinfo{author}{\bibfnamefont{C.~A.} \bibnamefont{Miller}},
  \bibinfo{journal}{Phys. Rev.} \textbf{\bibinfo{volume}{D70}},
  \bibinfo{pages}{117504} (\bibinfo{year}{2004}).

\bibitem[{\citenamefont{Georgi and Politzer}(1978)}]{GeorgiPolitzer78}
\bibinfo{author}{\bibfnamefont{H.}~\bibnamefont{Georgi}} \bibnamefont{and}
  \bibinfo{author}{\bibfnamefont{H.~D.} \bibnamefont{Politzer}},
  \bibinfo{journal}{Phys. Rev. Lett.} \textbf{\bibinfo{volume}{40}},
  \bibinfo{pages}{3} (\bibinfo{year}{1978}).

\bibitem[{\citenamefont{M\'{e}ndez}(1978)}]{Mendez78}
\bibinfo{author}{\bibfnamefont{A.}~\bibnamefont{M\'{e}ndez}},
  \bibinfo{journal}{Nucl. Phys.} \textbf{\bibinfo{volume}{B145}},
  \bibinfo{pages}{199 } (\bibinfo{year}{1978}).

\bibitem[{\citenamefont{Gl{\"u}ck et~al.}(1998)\citenamefont{Gl{\"u}ck, Reya,
  and Vogt}}]{Gluck:1998xa}
\bibinfo{author}{\bibfnamefont{M.}~\bibnamefont{Gl{\"u}ck}},
  \bibinfo{author}{\bibfnamefont{E.}~\bibnamefont{Reya}}, \bibnamefont{and}
  \bibinfo{author}{\bibfnamefont{A.}~\bibnamefont{Vogt}},
  \bibinfo{journal}{Eur. Phys. J.} \textbf{\bibinfo{volume}{C5}},
  \bibinfo{pages}{461} (\bibinfo{year}{1998}).

\bibitem[{\citenamefont{de~Florian
  et~al.}(2007{\natexlab{a}})\citenamefont{de~Florian, Sassot, and
  Stratmann}}]{deFlorian:2007aj}
\bibinfo{author}{\bibfnamefont{D.}~\bibnamefont{de~Florian}},
  \bibinfo{author}{\bibfnamefont{R.}~\bibnamefont{Sassot}}, \bibnamefont{and}
  \bibinfo{author}{\bibfnamefont{M.}~\bibnamefont{Stratmann}},
  \bibinfo{journal}{Phys. Rev.} \textbf{\bibinfo{volume}{D75}},
  \bibinfo{pages}{114010} (\bibinfo{year}{2007}{\natexlab{a}}).

\bibitem[{\citenamefont{Wandzura and Wilczek}(1977)}]{Wandzura:1977qf}
\bibinfo{author}{\bibfnamefont{S.}~\bibnamefont{Wandzura}} \bibnamefont{and}
  \bibinfo{author}{\bibfnamefont{F.}~\bibnamefont{Wilczek}},
  \bibinfo{journal}{Phys. Lett.} \textbf{\bibinfo{volume}{B72}},
  \bibinfo{pages}{195} (\bibinfo{year}{1977}).

\bibitem[{\citenamefont{Aubert et~al.}(1983)}]{Aubert:1983cz}
\bibinfo{author}{\bibfnamefont{J.~J.} \bibnamefont{Aubert}}
  \bibnamefont{et~al.} (\bibinfo{collaboration}{European Muon Collaboration}),
  \bibinfo{journal}{Phys. Lett.} \textbf{\bibinfo{volume}{B130}},
  \bibinfo{pages}{118} (\bibinfo{year}{1983}).

\bibitem[{\citenamefont{Arneodo et~al.}(1987)}]{Arneodo:1986cf}
\bibinfo{author}{\bibfnamefont{M.}~\bibnamefont{Arneodo}} \bibnamefont{et~al.}
  (\bibinfo{collaboration}{European Muon Collaboration}), \bibinfo{journal}{Z.
  Phys.} \textbf{\bibinfo{volume}{C34}}, \bibinfo{pages}{277}
  (\bibinfo{year}{1987}).

\bibitem[{\citenamefont{Breitweg et~al.}(2000)}]{Breitweg:2000qh}
\bibinfo{author}{\bibfnamefont{J.}~\bibnamefont{Breitweg}} \bibnamefont{et~al.}
  (\bibinfo{collaboration}{{\sc Zeus} Collaboration}), \bibinfo{journal}{Phys.
  Lett.} \textbf{\bibinfo{volume}{B481}}, \bibinfo{pages}{199}
  (\bibinfo{year}{2000}).

\bibitem[{\citenamefont{Adams et~al.}(1993)}]{Adams:1993hs}
\bibinfo{author}{\bibfnamefont{M.~R.} \bibnamefont{Adams}} \bibnamefont{et~al.}
  (\bibinfo{collaboration}{E665 Collaboration}), \bibinfo{journal}{Phys. Rev.}
  \textbf{\bibinfo{volume}{D48}}, \bibinfo{pages}{5057} (\bibinfo{year}{1993}).

\bibitem[{\citenamefont{Osipenko et~al.}(2009)}]{Osipenko:2008rv}
\bibinfo{author}{\bibfnamefont{M.}~\bibnamefont{Osipenko}} \bibnamefont{et~al.}
  (\bibinfo{collaboration}{{\sc Clas} Collaboration}), \bibinfo{journal}{Phys.
  Rev.} \textbf{\bibinfo{volume}{D80}}, \bibinfo{pages}{032004}
  (\bibinfo{year}{2009}).

\bibitem[{\citenamefont{Sbrizzai}(2011)}]{Sbrizzai}
\bibinfo{author}{\bibfnamefont{G.}~\bibnamefont{Sbrizzai}}
  (\bibinfo{collaboration}{{\sc Compass} Collaboration}), \bibinfo{journal}{J.
  Phys.: Conf. Ser.} \textbf{\bibinfo{volume}{295}}, \bibinfo{pages}{012043}
  (\bibinfo{year}{2011}).

\bibitem[{\citenamefont{Falciano et~al.}(1986)}]{Falciano:1986wk}
\bibinfo{author}{\bibfnamefont{S.}~\bibnamefont{Falciano}} \bibnamefont{et~al.}
  (\bibinfo{collaboration}{NA10 Collaboration}), \bibinfo{journal}{Z. Phys.}
  \textbf{\bibinfo{volume}{C31}}, \bibinfo{pages}{513} (\bibinfo{year}{1986}).

\bibitem[{\citenamefont{Guanziroli et~al.}(1988)}]{Guanziroli:1987rp}
\bibinfo{author}{\bibfnamefont{M.}~\bibnamefont{Guanziroli}}
  \bibnamefont{et~al.} (\bibinfo{collaboration}{NA10 Collaboration}),
  \bibinfo{journal}{Z. Phys.} \textbf{\bibinfo{volume}{C37}},
  \bibinfo{pages}{545} (\bibinfo{year}{1988}).

\bibitem[{\citenamefont{Conway et~al.}(1989)}]{Conway:1989fs}
\bibinfo{author}{\bibfnamefont{J.~S.} \bibnamefont{Conway}}
  \bibnamefont{et~al.}, \bibinfo{journal}{Phys. Rev.}
  \textbf{\bibinfo{volume}{D39}}, \bibinfo{pages}{92} (\bibinfo{year}{1989}).

\bibitem[{\citenamefont{Heinrich et~al.}(1991)}]{Heinrich:1991zm}
\bibinfo{author}{\bibfnamefont{J.~G.} \bibnamefont{Heinrich}}
  \bibnamefont{et~al.}, \bibinfo{journal}{Phys. Rev.}
  \textbf{\bibinfo{volume}{D44}}, \bibinfo{pages}{1909} (\bibinfo{year}{1991}).

\bibitem[{\citenamefont{Zhu et~al.}(2007)}]{Zhu:2006gx}
\bibinfo{author}{\bibfnamefont{L.~Y.} \bibnamefont{Zhu}} \bibnamefont{et~al.}
  (\bibinfo{collaboration}{FNAL-E866/NuSea Collaboration}),
  \bibinfo{journal}{Phys. Rev. Lett.} \textbf{\bibinfo{volume}{99}},
  \bibinfo{pages}{082301} (\bibinfo{year}{2007}).

\bibitem[{\citenamefont{Zhu et~al.}(2009)}]{Zhu:2008sj}
\bibinfo{author}{\bibfnamefont{L.~Y.} \bibnamefont{Zhu}} \bibnamefont{et~al.}
  (\bibinfo{collaboration}{FNAL E866/NuSea Collaboration}),
  \bibinfo{journal}{Phys. Rev. Lett.} \textbf{\bibinfo{volume}{102}},
  \bibinfo{pages}{182001} (\bibinfo{year}{2009}).

\bibitem[{\citenamefont{Lam and Tung}(1980)}]{Lam:1980uc}
\bibinfo{author}{\bibfnamefont{C.~S.} \bibnamefont{Lam}} \bibnamefont{and}
  \bibinfo{author}{\bibfnamefont{W.-K.} \bibnamefont{Tung}},
  \bibinfo{journal}{Phys. Rev.} \textbf{\bibinfo{volume}{D21}},
  \bibinfo{pages}{2712} (\bibinfo{year}{1980}).

\bibitem[{\citenamefont{Boer and Mulders}(2000)}]{Boer:1999si}
\bibinfo{author}{\bibfnamefont{D.}~\bibnamefont{Boer}} \bibnamefont{and}
  \bibinfo{author}{\bibfnamefont{P.~J.} \bibnamefont{Mulders}},
  \bibinfo{journal}{Nucl. Phys.} \textbf{\bibinfo{volume}{B569}},
  \bibinfo{pages}{505} (\bibinfo{year}{2000}).

\bibitem[{\citenamefont{Ackerstaff et~al.}(1998)}]{Ackerstaff:1998av}
\bibinfo{author}{\bibfnamefont{K.}~\bibnamefont{Ackerstaff}}
  \bibnamefont{et~al.} (\bibinfo{collaboration}{{\hermes} Collaboration}),
  \bibinfo{journal}{Nucl. Instrum. Meth.} \textbf{\bibinfo{volume}{A417}},
  \bibinfo{pages}{230} (\bibinfo{year}{1998}).

\bibitem[{\citenamefont{Fr{\"u}hwirth}(1987)}]{Fruhwirth:1987fm}
\bibinfo{author}{\bibfnamefont{R.}~\bibnamefont{Fr{\"u}hwirth}},
  \bibinfo{journal}{Nucl. Instrum. Meth.} \textbf{\bibinfo{volume}{A262}},
  \bibinfo{pages}{444} (\bibinfo{year}{1987}).

\bibitem[{\citenamefont{Akopov et~al.}(2002)}]{Akopov:2000qi}
\bibinfo{author}{\bibfnamefont{N.}~\bibnamefont{Akopov}} \bibnamefont{et~al.},
  \bibinfo{journal}{Nucl. Instrum. Meth.} \textbf{\bibinfo{volume}{A479}},
  \bibinfo{pages}{511} (\bibinfo{year}{2002}).

\bibitem[{\citenamefont{Jackson}(2005)}]{Jackson:2005eu}
\bibinfo{author}{\bibfnamefont{H.~E.} \bibnamefont{Jackson}}
  (\bibinfo{collaboration}{{\hermes\ Collaboration}}), \bibinfo{journal}{Nucl.
  Instrum. Meth.} \textbf{\bibinfo{volume}{A553}}, \bibinfo{pages}{205}
  (\bibinfo{year}{2005}).

\bibitem[{\citenamefont{Avakian et~al.}(1998)}]{Avakian:1998bz}
\bibinfo{author}{\bibfnamefont{H.}~\bibnamefont{Avakian}} \bibnamefont{et~al.},
  \bibinfo{journal}{Nucl. Instrum. Meth.} \textbf{\bibinfo{volume}{A417}},
  \bibinfo{pages}{69} (\bibinfo{year}{1998}).

\bibitem[{\citenamefont{Sj{\"o}strand et~al.}(2001)}]{PYTHIA6}
\bibinfo{author}{\bibfnamefont{T.}~\bibnamefont{Sj{\"o}strand}}
  \bibnamefont{et~al.}, \bibinfo{journal}{Comput. Phys. Commun.}
  \textbf{\bibinfo{volume}{135}}, \bibinfo{pages}{238} (\bibinfo{year}{2001}).

\bibitem[{\citenamefont{Sj{\"o}strand}(1994)}]{Sjostrand:1993yb}
\bibinfo{author}{\bibfnamefont{T.}~\bibnamefont{Sj{\"o}strand}},
  \bibinfo{journal}{Comput. Phys. Commun.} \textbf{\bibinfo{volume}{82}},
  \bibinfo{pages}{74} (\bibinfo{year}{1994}).

\bibitem[{\citenamefont{Airapetian
  et~al.}(2010{\natexlab{b}})}]{Airapetian:2010um}
\bibinfo{author}{\bibfnamefont{A.}~\bibnamefont{Airapetian}}
  \bibnamefont{et~al.} (\bibinfo{collaboration}{\hermes\ Collaboration}),
  \bibinfo{journal}{JHEP} \textbf{\bibinfo{volume}{1008}}, \bibinfo{pages}{130}
  (\bibinfo{year}{2010}{\natexlab{b}}).

\bibitem[{\citenamefont{Akushevich et~al.}(1998)\citenamefont{Akushevich,
  B{\"o}ttcher, and Ryckbosch}}]{radgen}
\bibinfo{author}{\bibfnamefont{I.}~\bibnamefont{Akushevich}},
  \bibinfo{author}{\bibfnamefont{H.}~\bibnamefont{B{\"o}ttcher}},
  \bibnamefont{and} \bibinfo{author}{\bibfnamefont{D.}~\bibnamefont{Ryckbosch}}
  (\bibinfo{year}{1998}), \eprint{hep-ph/9906408}.

\bibitem[{\citenamefont{Brun et~al.}(1978)\citenamefont{Brun, Hagelberg,
  Hansroul, and Lassalle}}]{geant}
\bibinfo{author}{\bibfnamefont{R.}~\bibnamefont{Brun}},
  \bibinfo{author}{\bibfnamefont{R.}~\bibnamefont{Hagelberg}},
  \bibinfo{author}{\bibfnamefont{M.}~\bibnamefont{Hansroul}}, \bibnamefont{and}
  \bibinfo{author}{\bibfnamefont{J.}~\bibnamefont{Lassalle}},
  \bibinfo{journal}{{\sc Cern} Report {\sc Cern}-DD-78-2-REV}
  (\bibinfo{year}{1978}).

\bibitem[{\citenamefont{Ingelman et~al.}(1997)\citenamefont{Ingelman, Edin, and
  Rathsman}}]{Ingelman:1996mq}
\bibinfo{author}{\bibfnamefont{G.}~\bibnamefont{Ingelman}},
  \bibinfo{author}{\bibfnamefont{A.}~\bibnamefont{Edin}}, \bibnamefont{and}
  \bibinfo{author}{\bibfnamefont{J.}~\bibnamefont{Rathsman}},
  \bibinfo{journal}{Comput. Phys. Commun.} \textbf{\bibinfo{volume}{101}},
  \bibinfo{pages}{108} (\bibinfo{year}{1997}).

\bibitem[{\citenamefont{Giordano}(2008)}]{GiordanoTr08}
\bibinfo{author}{\bibfnamefont{F.}~\bibnamefont{Giordano}}
  (\bibinfo{year}{2008}), \bibinfo{note}{proceedings of Second International
  Workshop on Transverse Polarisation Phenomena in Hard Processes (Transversity
  2008), Ferrara, Italy, 28 - 31 May 2008, World Scientific, 2008, p. 177}.

\bibitem[{Dat()}]{Database}
\bibinfo{note}{Durham HEP database, \url{http://durpdg.dur.ac.uk}; {\sc
  inSPIRE}, \url{http://inspirebeta.net/record/1111237/}; \\ mail-to:
  \href{mailto:management@hermes.desy.de}{management@hermes.desy.de}.}

\bibitem[{che()}]{cherrypicker}
\bibinfo{note}{\url{http://www-hermes.desy.de/cosnphi/}}.

\bibitem[{\citenamefont{Airapetian et~al.}()}]{dc19}
\bibinfo{author}{\bibfnamefont{A.}~\bibnamefont{Airapetian}}
  \bibnamefont{et~al.} (\bibinfo{collaboration}{\hermes\ Collaboration}),
  \bibinfo{note}{arXiv:1212.5407 [hep-ex]}.

\bibitem[{\citenamefont{de~Florian
  et~al.}(2007{\natexlab{b}})\citenamefont{de~Florian, Sassot, and
  Stratmann}}]{deFlorian:2007hc}
\bibinfo{author}{\bibfnamefont{D.}~\bibnamefont{de~Florian}},
  \bibinfo{author}{\bibfnamefont{R.}~\bibnamefont{Sassot}}, \bibnamefont{and}
  \bibinfo{author}{\bibfnamefont{M.}~\bibnamefont{Stratmann}},
  \bibinfo{journal}{Phys. Rev.} \textbf{\bibinfo{volume}{D76}},
  \bibinfo{pages}{074033} (\bibinfo{year}{2007}{\natexlab{b}}).

\bibitem[{\citenamefont{Schnell}(2007)}]{Schnell:2007}
\bibinfo{author}{\bibfnamefont{G.}~\bibnamefont{Schnell}}
  (\bibinfo{collaboration}{{\hermes\ Collaboration}}) (\bibinfo{year}{2007}),
  \bibinfo{note}{talk delivered at `Transverse momentum, spin, and position
  distributions of partons in hadrons` (ECT* 2007), Trento, Italy, 11 - 15 Jun
  2007.}

\bibitem[{\citenamefont{Airapetian et~al.}(2005)}]{Aut2005}
\bibinfo{author}{\bibfnamefont{A.}~\bibnamefont{Airapetian}}
  \bibnamefont{et~al.} (\bibinfo{collaboration}{\hermes\ Collaboration}),
  \bibinfo{journal}{Phys. Rev. Lett.} \textbf{\bibinfo{volume}{94}},
  \bibinfo{pages}{012002} (\bibinfo{year}{2005}).

\bibitem[{\citenamefont{Anselmino
  et~al.}(2007{\natexlab{a}})}]{Anselmino:2007fs}
\bibinfo{author}{\bibfnamefont{M.}~\bibnamefont{Anselmino}}
  \bibnamefont{et~al.}, \bibinfo{journal}{Phys. Rev.}
  \textbf{\bibinfo{volume}{D75}}, \bibinfo{pages}{054032}
  (\bibinfo{year}{2007}{\natexlab{a}}).

\bibitem[{\citenamefont{Airapetian
  et~al.}(2010{\natexlab{c}})}]{Airapetian:2010ds}
\bibinfo{author}{\bibfnamefont{A.}~\bibnamefont{Airapetian}}
  \bibnamefont{et~al.} (\bibinfo{collaboration}{\hermes\ Collaboration}),
  \bibinfo{journal}{Phys. Lett.} \textbf{\bibinfo{volume}{B693}},
  \bibinfo{pages}{11} (\bibinfo{year}{2010}{\natexlab{c}}).

\bibitem[{\citenamefont{Alexakhin et~al.}(2005)}]{Alexakhin:2005iw}
\bibinfo{author}{\bibfnamefont{V.~Y.} \bibnamefont{Alexakhin}}
  \bibnamefont{et~al.} (\bibinfo{collaboration}{{\compass\ Collaboration}}),
  \bibinfo{journal}{Phys. Rev. Lett.} \textbf{\bibinfo{volume}{94}},
  \bibinfo{pages}{202002} (\bibinfo{year}{2005}).

\bibitem[{\citenamefont{Abe et~al.}(2006)}]{Abe:2005zx}
\bibinfo{author}{\bibfnamefont{K.}~\bibnamefont{Abe}} \bibnamefont{et~al.}
  (\bibinfo{collaboration}{\belle\ Collaboration}), \bibinfo{journal}{Phys.
  Rev. Lett.} \textbf{\bibinfo{volume}{96}}, \bibinfo{pages}{232002}
  (\bibinfo{year}{2006}).

\bibitem[{\citenamefont{Lamb}(2010)}]{RebeccaThesis}
\bibinfo{author}{\bibfnamefont{R.}~\bibnamefont{Lamb}}, Ph.D. thesis,
  \bibinfo{school}{University of Illinois at Urbana-Champaign}
  (\bibinfo{year}{2010}), \bibinfo{note}{{DESY-THESIS-2010-035}}.

\bibitem[{\citenamefont{Barone et~al.}(2010)\citenamefont{Barone, Melis, and
  Prokudin}}]{Barone:2009hw}
\bibinfo{author}{\bibfnamefont{V.}~\bibnamefont{Barone}},
  \bibinfo{author}{\bibfnamefont{S.}~\bibnamefont{Melis}}, \bibnamefont{and}
  \bibinfo{author}{\bibfnamefont{A.}~\bibnamefont{Prokudin}},
  \bibinfo{journal}{Phys. Rev.} \textbf{\bibinfo{volume}{D81}},
  \bibinfo{pages}{114026} (\bibinfo{year}{2010}).

\bibitem[{\citenamefont{Burkardt and Hannafious}(2008)}]{Burkardt:2007xm}
\bibinfo{author}{\bibfnamefont{M.}~\bibnamefont{Burkardt}} \bibnamefont{and}
  \bibinfo{author}{\bibfnamefont{B.}~\bibnamefont{Hannafious}},
  \bibinfo{journal}{Phys. Lett.} \textbf{\bibinfo{volume}{B658}},
  \bibinfo{pages}{130} (\bibinfo{year}{2008}).

\bibitem[{\citenamefont{Burkardt}(2005)}]{Burkardt:2005hp}
\bibinfo{author}{\bibfnamefont{M.}~\bibnamefont{Burkardt}},
  \bibinfo{journal}{Phys. Rev.} \textbf{\bibinfo{volume}{D72}},
  \bibinfo{pages}{094020} (\bibinfo{year}{2005}).

\bibitem[{\citenamefont{Gamberg et~al.}(2003)\citenamefont{Gamberg, Goldstein,
  and Oganessyan}}]{Gamberg:2003ey}
\bibinfo{author}{\bibfnamefont{L.~P.} \bibnamefont{Gamberg}},
  \bibinfo{author}{\bibfnamefont{G.~R.} \bibnamefont{Goldstein}},
  \bibnamefont{and} \bibinfo{author}{\bibfnamefont{K.~A.}
  \bibnamefont{Oganessyan}}, \bibinfo{journal}{Phys. Rev.}
  \textbf{\bibinfo{volume}{D67}}, \bibinfo{pages}{071504}
  (\bibinfo{year}{2003}).

\bibitem[{\citenamefont{Gamberg et~al.}(2008)\citenamefont{Gamberg, Goldstein,
  and Schlegel}}]{Gamberg:2007wm}
\bibinfo{author}{\bibfnamefont{L.~P.} \bibnamefont{Gamberg}},
  \bibinfo{author}{\bibfnamefont{G.~R.} \bibnamefont{Goldstein}},
  \bibnamefont{and} \bibinfo{author}{\bibfnamefont{M.}~\bibnamefont{Schlegel}},
  \bibinfo{journal}{Phys. Rev.} \textbf{\bibinfo{volume}{D77}},
  \bibinfo{pages}{094016} (\bibinfo{year}{2008}).

\bibitem[{\citenamefont{Barone et~al.}(2008)\citenamefont{Barone, Prokudin, and
  Ma}}]{Barone:2008tn}
\bibinfo{author}{\bibfnamefont{V.}~\bibnamefont{Barone}},
  \bibinfo{author}{\bibfnamefont{A.}~\bibnamefont{Prokudin}}, \bibnamefont{and}
  \bibinfo{author}{\bibfnamefont{B.-Q.} \bibnamefont{Ma}},
  \bibinfo{journal}{Phys. Rev.} \textbf{\bibinfo{volume}{D78}},
  \bibinfo{pages}{045022} (\bibinfo{year}{2008}).

\bibitem[{\citenamefont{Zhang et~al.}(2008)\citenamefont{Zhang, Lu, Ma, and
  Schmidt}}]{Zhang:2008ez}
\bibinfo{author}{\bibfnamefont{B.}~\bibnamefont{Zhang}},
  \bibinfo{author}{\bibfnamefont{Z.}~\bibnamefont{Lu}},
  \bibinfo{author}{\bibfnamefont{B.-Q.} \bibnamefont{Ma}}, \bibnamefont{and}
  \bibinfo{author}{\bibfnamefont{I.}~\bibnamefont{Schmidt}},
  \bibinfo{journal}{Phys. Rev.} \textbf{\bibinfo{volume}{D78}},
  \bibinfo{pages}{034035} (\bibinfo{year}{2008}).

\bibitem[{\citenamefont{Giordano and Lamb}(2009)}]{Giordano:2009hi}
\bibinfo{author}{\bibfnamefont{F.}~\bibnamefont{Giordano}} \bibnamefont{and}
  \bibinfo{author}{\bibfnamefont{R.}~\bibnamefont{Lamb}}
  (\bibinfo{collaboration}{\hermes\ Collaboration}), \bibinfo{journal}{AIP
  Conf. Proc.} \textbf{\bibinfo{volume}{1149}}, \bibinfo{pages}{423}
  (\bibinfo{year}{2009}).

\bibitem[{\citenamefont{Anselmino
  et~al.}(2005{\natexlab{a}})}]{Anselmino:2005ea}
\bibinfo{author}{\bibfnamefont{M.}~\bibnamefont{Anselmino}}
  \bibnamefont{et~al.}, \bibinfo{journal}{Phys. Rev.}
  \textbf{\bibinfo{volume}{D72}}, \bibinfo{pages}{094007}
  (\bibinfo{year}{2005}{\natexlab{a}}).

\bibitem[{\citenamefont{Anselmino
  et~al.}(2005{\natexlab{b}})}]{Anselmino:2005nn}
\bibinfo{author}{\bibfnamefont{M.}~\bibnamefont{Anselmino}}
  \bibnamefont{et~al.}, \bibinfo{journal}{Phys. Rev.}
  \textbf{\bibinfo{volume}{D71}}, \bibinfo{pages}{074006}
  (\bibinfo{year}{2005}{\natexlab{b}}).

\bibitem[{\citenamefont{Aybat and Rogers}(2011)}]{Aybat:2011zv}
\bibinfo{author}{\bibfnamefont{S.~M.} \bibnamefont{Aybat}} \bibnamefont{and}
  \bibinfo{author}{\bibfnamefont{T.~C.} \bibnamefont{Rogers}},
  \bibinfo{journal}{Phys. Rev.} \textbf{\bibinfo{volume}{D83}},
  \bibinfo{pages}{114042} (\bibinfo{year}{2011}).

\bibitem[{\citenamefont{Anselmino
  et~al.}(2007{\natexlab{b}})\citenamefont{Anselmino, Boglione, Prokudin, and
  Turk}}]{Anselmino:2006rv}
\bibinfo{author}{\bibfnamefont{M.}~\bibnamefont{Anselmino}},
  \bibinfo{author}{\bibfnamefont{M.}~\bibnamefont{Boglione}},
  \bibinfo{author}{\bibfnamefont{A.}~\bibnamefont{Prokudin}}, \bibnamefont{and}
  \bibinfo{author}{\bibfnamefont{C.}~\bibnamefont{Turk}},
  \bibinfo{journal}{Eur. Phys. J.} \textbf{\bibinfo{volume}{A31}},
  \bibinfo{pages}{373} (\bibinfo{year}{2007}{\natexlab{b}}).

\bibitem[{\citenamefont{Artru et~al.}(1997)\citenamefont{Artru, Czyzewski, and
  Yabuki}}]{Artru:1997bh}
\bibinfo{author}{\bibfnamefont{X.}~\bibnamefont{Artru}},
  \bibinfo{author}{\bibfnamefont{J.}~\bibnamefont{Czyzewski}},
  \bibnamefont{and} \bibinfo{author}{\bibfnamefont{H.}~\bibnamefont{Yabuki}},
  \bibinfo{journal}{Z. Phys.} \textbf{\bibinfo{volume}{C73}},
  \bibinfo{pages}{527} (\bibinfo{year}{1997}).

\bibitem[{\citenamefont{Airapetian et~al.}(2008)}]{Airapetian2008qf}
\bibinfo{author}{\bibfnamefont{A.}~\bibnamefont{Airapetian}}
  \bibnamefont{et~al.} (\bibinfo{collaboration}{\hermes\ Collaboration}),
  \bibinfo{journal}{Phys. Lett.} \textbf{\bibinfo{volume}{B666}},
  \bibinfo{pages}{446} (\bibinfo{year}{2008}).

\bibitem[{\citenamefont{Cisbani}(1999)}]{Cisbani1999366}
\bibinfo{author}{\bibfnamefont{E.}~\bibnamefont{Cisbani}},
  \bibinfo{journal}{Nuclear Physics B - Proceedings Supplements}
  \textbf{\bibinfo{volume}{78}}, \bibinfo{pages}{366 } (\bibinfo{year}{1999}).

\end{thebibliography}
